\documentclass[10pt]{article}
\pdfoutput=1
\usepackage{amssymb,amsmath,mathrsfs}
\usepackage{graphicx,rotate,subfigure}
\usepackage{cite}
\usepackage[colorlinks=true,
            linkcolor=red,
            urlcolor=blue,
            citecolor=blue]{hyperref}

\usepackage{color}
\long\def\rpl#1!!#2!!{\textcolor{red}{#1} \textcolor{blue}{#2}}

\def \order(#1){{\cal O} \left(#1 \right)}

\evensidemargin=\oddsidemargin
\def\Eqn#1{Eq.\ (\ref{#1})}

\def\m{\scriptstyle}
\def\l{\lambda}
\allowdisplaybreaks
\begin{document}

\begin{flushright}
SINP/TNP/2014/22 \\
HRI-RECAPP-2014-006
\end{flushright}

\begin{center}
{\Large \bf Analysis of an extended scalar sector with $S_3$ symmetry} \\
\vspace*{1cm}  {\sf
 Dipankar Das$^\dagger$\footnote{d.das@saha.ac.in}, ~ Ujjal Kumar Dey$^\ddagger$\footnote{ujjaldey@hri.res.in}}
\\
\vspace{10pt} {\small } $^\dagger${\em Saha Institute of Nuclear
    Physics, 1/AF Bidhan Nagar, Kolkata 700064, India}

\vspace{10pt} {\small } $^\ddagger${\em Harish-Chandra Research Institute, Chhatnag Road, Jhunsi, Allahabad 211019, India}
\normalsize
\end{center}

\begin{abstract}
We investigate the scalar potential of a general $S_3$-symmetric three-Higgs-doublet model. The outcome of our analysis does not depend on the fermionic sector of the model. We identify a decoupling limit for the scalar spectrum of this scenario. In view of the recent LHC Higgs data, we show our numerical results only in the decoupling limit. Unitarity and stability of the scalar potential demand that many new scalars must be lurking below 1 TeV. We provide numerical predictions for $h\to \gamma \gamma$ and $h\to Z \gamma$ signal strengths which can be used to falsify the theory.

\end{abstract}

\bigskip

\section{Introduction}
The newly observed boson at the Large Hadron Collider (LHC)  \cite{Aad:2012tfa,Chatrchyan:2012ufa} fits very well to the description of the Higgs scalar in the Standard Model (SM). The SM relies on the minimal choice that a single Higgs doublet provides masses to all particles. But unexplained phenomena like neutrino masses and existence dark matter motivate us to contemplate other avenues beyond the SM (BSM). Majority of these BSM scenarios extend the SM Higgs sector predicting a richer scalar spectrum. One of them $-$ the $S_3$ flavor model $-$ stems from an effort to answer the aesthetic question as to why there are precisely three fermion generations \cite{Kubo:2003pd}. Keeping the fermions in appropriate $S_3$-multiplets, it is possible to reproduce all the measured parameters of the CKM and PMNS matrices as well as make testable predictions for the unknown parameters of the PMNS matrix \cite{ Koide:1999mx,Harrison:2003aw,Kubo:2003iw,Teshima:2005bk,Koide:2006vs,Chen:2007zj,Mondragon:2007af,Jora:2009gz,Xing:2010iu,Kaneko:2010rx,Zhou:2011nu,Teshima:2011wg,Dev:2011qy,Dev:2012ns,Meloni:2012ci, Dias:2012bh,Siyeon:2012zu,Canales:2012dr, Canales:2013cga,Benaoum:2013ji,Hernandez:2014vta}. But one needs at least three Higgs doublets to achieve this goal \cite{Kubo:2003iw}. However, the $S_3$ invariant scalar potential contains some new parameters which are difficult to constrain phenomenologically. Although some lower bounds on the additional scalar masses can be placed from the Higgs mediated flavor changing neutral current (FCNC) processes \cite{Ma:2013zca}, these bounds rely heavily on the Yukawa structure of the model. In this article we will present some new  bounds on the physical scalar masses which do not depend on the parameters of the Yukawa sector. To achieve this, we will employ the prescription of tree unitarity which is known to be able to set upper limits on different scalar masses \cite{Lee:1977eg}. Although various aspects of the $S_3$ scalar potential have been discussed in the literature \cite{Pakvasa:1977in,Kubo:2004ps}, to the best of our knowledge, this is the first attempt to derive the exact unitarity constraints on the quartic couplings in the $S_3$ invariant three-Higgs-doublet model (S3HDM) scalar potential. We also identify a \emph{decoupling limit} in the context of S3HDM where a CP-even Higgs with SM-like properties can be obtained. Since the recent LHC Higgs data seem to increasingly leaned towards the SM expectations, our numerical analysis will be restricted to this limit.  
 
The paper is organized as follows~: in Section~\ref{s:potential} we discuss the scalar potential and derive necessary conditions for the potential to be bounded from below. In Section~\ref{s:physeig} we minimize the potential and calculate the physical scalar masses. In this section we also figure out a decoupling limit in which one neutral CP-even physical scalar behaves exactly like the SM Higgs. In Section~\ref{s:unitarity} we derive the exact constraints arising from the considerations of tree level unitarity and use them to constrain the nonstandard scalar masses. In Section~\ref{s:decay} we quantitatively investigate the effect of the charged scalar induced loops on $h\to \gamma \gamma$ and $h\to Z \gamma$ signal strengths. Finally, we summarize our findings in Section~\ref{s:concl}.

\section{The scalar potential} \label{s:potential}
%
$S_3$ is the permutation group involving three objects, $\{\phi_a, \phi_b, \phi_c\}$. The three dimensional representation of $S_3$ is not an irreducible one simply because we can easily construct a linear combination of the elements, $\phi_a+\phi_b+\phi_c$, which remains unaltered under the permutation of the indices. We choose to decompose the three dimensional representation into a singlet and doublet as follows~:
\begin{subequations}
\begin{eqnarray}
{\bf 1}~: && ~~ \phi_3=\frac{1}{\sqrt{3}}(\phi_a+\phi_b+\phi_c) \,, \\
{\bf 2}~: && ~ \begin{pmatrix}\phi_1 \\ \phi_2 \end{pmatrix} = 
 \begin{pmatrix} \frac{1}{\sqrt{2}}(\phi_a-\phi_b) \\ \frac{1}{\sqrt{6}}(\phi_a+\phi_b-2\phi_c) \end{pmatrix}\,.
\end{eqnarray}
\end{subequations}
The elements of $S_3$ for this particular doublet representation are given by~:
\begin{eqnarray}
 \begin{pmatrix}\cos\theta & \sin\theta \\ -\sin\theta & \cos\theta \end{pmatrix} \,,~~
 \begin{pmatrix}\cos\theta & \sin\theta \\ \sin\theta & -\cos\theta \end{pmatrix}\,, ~~
 {\rm for}~~ \left(\theta=0,\pm \frac{2\pi}{3}\right) \,.
\end{eqnarray}
The most general renormalizable potential invariant under $S_3$ can be written in terms of $\phi_3$, $\phi_1$ and $\phi_2$ as follows~\cite{Pakvasa:1977in,Kubo:2004ps,Koide:2005ep,Teshima:2012cg,Machado:2012ed}:
\begin{subequations}
\begin{eqnarray}
V(\phi)&=& V_2(\phi) +V_4(\phi) \,, \\
{\rm where,}~~ V_2(\phi) &=& \mu_1^2(\phi_1^\dagger\phi_1+\phi_2^\dagger\phi_2)+ \mu_3^2\phi_3^\dagger\phi_3 \,, \\
V_4(\phi)&=& \lambda_1 (\phi_1^\dagger\phi_1+\phi_2^\dagger\phi_2)^2 +\lambda_2 (\phi_1^\dagger\phi_2 -\phi_2^\dagger\phi_1)^2 +\lambda_3 \left\{(\phi_1^\dagger\phi_2+\phi_2^\dagger\phi_1)^2 +(\phi_1^\dagger\phi_1-\phi_2^\dagger\phi_2) ^2\right\} \nonumber \\
&& +\lambda_4 \left\{(\phi_3^\dagger\phi_1)(\phi_1^\dagger\phi_2+\phi_2^\dagger\phi_1) +(\phi_3^\dagger\phi_2)(\phi_1^\dagger\phi_1-\phi_2^\dagger\phi_2) + {\rm h. c.}\right\} \nonumber \\
&& +\lambda_5(\phi_3^\dagger\phi_3)(\phi_1^\dagger\phi_1+\phi_2^\dagger\phi_2) + \lambda_6 \left\{(\phi_3^\dagger\phi_1)(\phi_1^\dagger\phi_3)+(\phi_3^\dagger\phi_2)(\phi_2^\dagger\phi_3)\right\} \nonumber \\
&& +\lambda_7 \left\{(\phi_3^\dagger\phi_1)(\phi_3^\dagger\phi_1) + (\phi_3^\dagger\phi_2)(\phi_3^\dagger\phi_2) +{\rm h. c.}\right\} +\lambda_8(\phi_3^\dagger\phi_3)^2 \,.
\label{quartic}
\end{eqnarray}
\label{potential}
\end{subequations}
In general $\lambda_4$ and $\lambda_7$ can be complex, but we assume them to be real so that CP symmetry is not broken explicitly. For the stability of the vacuum in the asymptotic limit we impose the requirement that there should be no direction in the field space along which the potential becomes infinitely negative. The necessary and sufficient conditions for this is well known in the context of two Higgs-doublet models (2HDMs) \cite{Gunion:2002zf}. For the potential of \Eqn{potential}, a 2HDM equivalent situation arise if one of the doublets is made identically zero. Then it is quite straightforward to find the following {\em necessary} conditions for the global stability in the asymptotic limit~:
\begin{subequations}
\begin{eqnarray}
\lambda_1 &>& 0 \,, \\
\lambda_8 &>& 0 \,, \\
\lambda_1+\lambda_3 &>& 0 \,, \\
2\lambda_1 +(\lambda_3-\lambda_2) &>& |\lambda_2+\lambda_3| \,, \\
\lambda_5 +2\sqrt{\lambda_8(\lambda_1+\lambda_3)} &>& 0 \,, \\
\lambda_5+\lambda_6+ 2\sqrt{\lambda_8(\lambda_1+\lambda_3)} &>& 2|\lambda_7| \,, \\
\lambda_1+\lambda_3+\lambda_5+\lambda_6+2\lambda_7+\lambda_8 &>& 2|\lambda_4| \,.
\end{eqnarray}
\label{stability}
\end{subequations}
To avoid confusion, we wish to mention that an equivalent doublet representation,
\begin{eqnarray}
\begin{pmatrix}\chi_1 \\ \chi_2 \end{pmatrix} =  \frac{1}{\sqrt{2}}\begin{pmatrix}i & 1 \\ -i & 1 \end{pmatrix}
 \begin{pmatrix} \phi_1 \\ \phi_2 \end{pmatrix}\,,
\end{eqnarray}
has also been used in the literature. In terms of this new doublet, the quartic part of the scalar potential is written as~\cite{Bhattacharyya:2010hp,Bhattacharyya:2012ze,Chen:2004rr}:
\begin{eqnarray}
V_4 &=& \frac{\beta_1}{2}\left(\chi_1^\dagger\chi_1+\chi_2^\dagger\chi_2 \right)^2 +\frac{\beta_2}{2}\left(\chi_1^\dagger\chi_1- \chi_2^\dagger\chi_2 \right)^2 +\beta_3(\chi_1^\dagger\chi_2)(\chi_2^\dagger\chi_1) +\frac{\beta_4}{2}(\phi_3^\dagger\phi_3)^2 \nonumber \\
&& +\beta_5(\phi_3^\dagger\phi_3)(\chi_1^\dagger\chi_1+\chi_2^\dagger\chi_2 ) +\beta_6\phi_3^\dagger (\chi_1\chi_1^\dagger +\chi_2\chi_2^\dagger )\phi_3+ \beta_7\left\{(\phi_3^\dagger\chi_1)(\phi_3^\dagger\chi_2)+{\rm h.c.}\right\} \nonumber \\
&& +\beta_8\left\{\phi_3^\dagger(\chi_1\chi_2^\dagger\chi_1+\chi_2\chi_1^\dagger\chi_2) + {\rm h.c.}\right\} \,.
\label{alternative}
\end{eqnarray}
It is easy to verify that the parameters of \Eqn{alternative} are related to the parameters of \Eqn{quartic} in the following way~:
\begin{eqnarray}
\beta_1 = 2\lambda_1~; ~~\beta_2=-2\lambda_2~; ~~\beta_3=4\lambda_3~; ~~ \beta_4=2\lambda_8~; ~~ \beta_5=\lambda_5~; ~~ \beta_6=\lambda_6~;~~ \beta_7=2\lambda_7~;~~ \beta_8=-\sqrt{2}\lambda_4 \,.
\end{eqnarray}
This mapping can be used to translate the constraints on $\lambda$s into constraints on $\beta$s.
In this paper we opt to work with the parametrization of \Eqn{potential}.

\section{Physical eigenstates} \label{s:physeig}
We represent the scalar doublets in the following way~:
\begin{eqnarray}
\phi_k= \begin{pmatrix} w_k^+ \\ \frac{1}{\sqrt{2}}(v_k+h_k+iz_k) \end{pmatrix} ~~~~{\rm for}~k=1,~2,~3\,.
\label{e:doub}
\end{eqnarray}
We shall assume that CP symmetry is not spontaneously broken and so the vacuum expectation values (vevs) are taken to be real. They also satisfy the usual vev relation~: $v=\sqrt{v_1^2+v_2^2+v_3^2}=$ 246 GeV. The minimization conditions for the scalar potential of \Eqn{potential} reads~:
\begin{subequations}
\begin{eqnarray}
\mu_1^2 &=& -2\lambda_1(v_1^2+v_2^2)-2\lambda_3(v_1^2+v_2^2)-v_3\{6\lambda_4v_2 +(\lambda_5+\lambda_6+2\lambda_7)v_3\} \,, \label{mu11}\\
\mu_1^2 &=& -2\lambda_1(v_1^2+v_2^2)-2\lambda_3(v_1^2+v_2^2) -\frac{3v_3}{v_2}\lambda_4(v_1^2-v_2^2) - (\lambda_5+\lambda_6+2\lambda_7)v_3^2 \,, \label{mu22}\\
\mu_3^2 &=& \lambda_4\frac{v_2}{v_3}(v_2^2-v_1^2) -(\lambda_5+\lambda_6+2\lambda_7)(v_1^2+v_2^2)-2\lambda_8 v_3^2 \,.
\end{eqnarray}
\label{minimization}
\end{subequations}
For the self-consistency of Eqs.~(\ref{mu11}) and (\ref{mu22}), two possible scenarios arise\footnote{Another possibility, $v_3=0$, while mathematically consistent, is unattractive. This is because, in some $S_3$ structure of the Yukawa sector, the $S_3$-singlet fermion generation will the remain massless.}~:
\begin{subequations}
\begin{eqnarray}
\lambda_4 &=& 0 \,, \\
{\rm or,}~~v_1 &=& \sqrt{3}v_2 \,.
\end{eqnarray}
\label{consistency}
\end{subequations}
In the following subsections we shall discuss each of the above scenarios separately.

\subsection{Case-I ($\lambda_4=0$)}
Since CP symmetry is assumed to be exact in the scalar potential, the neutral physical states will be eigenstates of CP too. We find that the mass-squared matrices in the scalar($M_S^2$), pseudoscalar($M_P^2$) and charged($M_C^2$) sectors are  simultaneously block diagonalizable by the following matrix~:
\begin{eqnarray}
X= \begin{pmatrix}
\cos\gamma & -\sin\gamma & 0 \\ \sin\gamma & \cos\gamma & 0 \\ 0 & 0 & 1
\end{pmatrix} ~~~{\rm with} ~~\tan\gamma=\frac{v_1}{v_2} \,.
\end{eqnarray}
For the charged mass matrix, we obtain~:
\begin{eqnarray}
XM_C^2X^T= \begin{pmatrix}
m_{1+}^2 & 0 & 0 \\ 0 & -\frac{1}{2}v_3^2(\lambda_6+2\lambda_7) & \frac{1}{2}v_3\sqrt{v_1^2+v_2^2} (\lambda_6+2\lambda_7) \\ 0& \frac{1}{2}v_3\sqrt{v_1^2+v_2^2}(\lambda_6+2\lambda_7) & -\frac{1}{2}(v_1^2+v_2^2) (\lambda_6+2\lambda_7)
\end{pmatrix} \,,
\end{eqnarray}
where, one of the charged Higgs ($H_1^+$) with mass $m_{1+}$ is defined as~:
\begin{subequations}
\begin{eqnarray}
H_1^+ &=& \cos\gamma ~w_1^+ -\sin\gamma ~w_2^+ \,, \\
m^2_{1+} &=& -\left\{2\lambda_3\sin^2\beta +\frac{1}{2}(\lambda_6+2\lambda_7)\cos^2\beta \right\}v^2 \,, \\
{\rm with,} ~~\tan\beta &=& \frac{\sqrt{v_1^2+v_2^2}}{v_3} \,.
\label{tanb}
\end{eqnarray}
\end{subequations}
The second charged Higgs ($H_2^+$) along with the massless Goldstone ($\omega^+$), which will appear as the longitudinal component of the $W$-boson, can be obtained by diagonalizing the remaining $2\times 2$ block~:
\begin{eqnarray}
\begin{pmatrix}
H_2^+ \\ \omega^+
\end{pmatrix}= \begin{pmatrix}
\cos\beta & -\sin\beta \\ \sin\beta & \cos\beta
\end{pmatrix} \begin{pmatrix}
w_2'^+ \\ w_3^+
\end{pmatrix} ~~~{\rm with,} ~~ w_2'^+ = \sin\gamma~w_1^+ +\cos\gamma~w_2^+ \,.
\end{eqnarray}
The mass of the second charged Higgs is given by~:
\begin{eqnarray}
m_{2+}^2 = -\frac{1}{2}(\lambda_6+2\lambda_7) v^2 \,.
\end{eqnarray}
Similar considerations for the pseudoscalar part gives~:
\begin{eqnarray}
XM_P^2X^T= \begin{pmatrix}
\frac{1}{2} m_{A1}^2 & 0 & 0 \\ 0 & -v_3^2\lambda_7 & v_3\sqrt{v_1^2+v_2^2}\lambda_7 \\ 0& v_3\sqrt{v_1^2+v_2^2}\lambda_7 & -(v_1^2+v_2^2)\lambda_7
\end{pmatrix} \,,
\end{eqnarray}
where, the pseudoscalar state ($A_1$) with mass eigenvalue $m_{A1}$ is defined as~:
\begin{subequations}
\begin{eqnarray}
A_1 &=& \cos\gamma ~z_1 -\sin\gamma ~z_2 \,, \\
m^2_{A1} &=& -2\left\{(\lambda_2+\lambda_3)\sin^2\beta +\lambda_7\cos^2\beta \right\}v^2 \,, 
\end{eqnarray}
\end{subequations}
where, $\tan\beta$ has already been defined in \Eqn{tanb}. Similar to the charged part, here also the second pseudoscalar ($A_2$) along with the massless Goldstone ($\zeta$) can be obtained as follows~:
\begin{subequations}
\begin{eqnarray}
\begin{pmatrix}
A_2 \\ \zeta
\end{pmatrix}&=& \begin{pmatrix}
\cos\beta & -\sin\beta \\ \sin\beta & \cos\beta
\end{pmatrix} \begin{pmatrix}
z_2' \\ z_3
\end{pmatrix} ~~~{\rm with,} ~~ z_2' = \sin\gamma~z_1 +\cos\gamma~z_2 \,, \\
{\rm and,} ~~~ m_{A2}^2 &=& -2\lambda_7 v^2 \,.
\end{eqnarray}
\end{subequations}
Finally, for the CP-even part we have~:
\begin{subequations}
\begin{eqnarray}
XM_S^2X^T &=& \begin{pmatrix}
0 & 0 & 0 \\ 0 & A'_S & -B'_S \\ 0& -B'_S &  C'_S
\end{pmatrix} \,, \\
{\rm where,} ~~~ A'_S &=& (\lambda_1+\lambda_3)(v_1^2+v_2^2) \,, \\
B'_S &=& -\frac{1}{2}v_3\sqrt{v_1^2+v_2^2}(\lambda_5+\lambda_6+2\lambda_7) \,, \\
C'_S &=& \lambda_8 v_3^2 \,.
\end{eqnarray}
\end{subequations}
The massless state ($h^0$), as also noted in \cite{Beltran:2009zz}, is given by~:
\begin{eqnarray}
h^0 &=& \cos\gamma ~h_1 -\sin\gamma ~h_2 \,.
\label{h0}
\end{eqnarray}
But we wish to add here that the appearance of a massless scalar is not surprising. One can easily verify that the potential of \Eqn{potential} has the following $SO(2)$ symmetry for $\lambda_4=0$~:
\begin{eqnarray}
\begin{pmatrix}
\phi_1' \\ \phi_2'
\end{pmatrix}&=& \begin{pmatrix}
\cos\theta & -\sin\theta \\ \sin\theta & \cos\theta
\end{pmatrix} \begin{pmatrix}
\phi_1 \\ \phi_2
\end{pmatrix}
\end{eqnarray}
Since $SO(2)$ is a continuous symmetry isomorphic to $U(1)$, a massless physical state is expected. Other two physical scalars are obtained as follows~:
\begin{subequations}
\begin{eqnarray}
\begin{pmatrix}h \\ H \end{pmatrix} &=& \begin{pmatrix}
\cos\alpha & -\sin\alpha \\ \sin\alpha & \cos\alpha
\end{pmatrix} \begin{pmatrix}
h_2' \\ h_3
\end{pmatrix} ~~~{\rm with,} ~~ h_2' = \sin\gamma~h_1 +\cos\gamma~h_2 \,, \\
{\rm and,}~~~ \tan2\alpha &=&\frac{2B'_S}{A'_S-C'_S} \,.
\end{eqnarray}
\end{subequations}
We assume $H$ and $h$ to be the heavier and lighter CP-even mass eigenstates respectively, with the following eigenvalues~:
\begin{subequations}
\begin{eqnarray}
m_H^2 &=& (A'_S+C'_S)+\sqrt{(A'_S-C'_S)^2+4B_S^{'2}} \,, \\
m_h^2 &=& (A'_S+C'_S)-\sqrt{(A'_S-C'_S)^2+4B_S^{'2}} \,.
\end{eqnarray}
\end{subequations}
At this stage, it is worth noting that we can define two intermediate scalar states, $H^0$ and $R$, as
\begin{eqnarray}
\begin{pmatrix}R \\ H^0 \end{pmatrix} &=& \begin{pmatrix}
\cos\beta & -\sin\beta \\ \sin\beta & \cos\beta
\end{pmatrix} \begin{pmatrix}
h_2' \\ h_3
\end{pmatrix} \,,
\label{H0R}
\end{eqnarray}
with the property that $H^0$ has the exact SM couplings with the vector boson pairs and fermions. $H^0$ does not take part in the flavor changing processes as well. Of course, $H^0$ and $R$ are not the physical eigenstates in general but are related to them in the following way~:
\begin{subequations}
\begin{eqnarray}
h &=& \cos(\beta-\alpha)R +\sin(\beta-\alpha) H^0 \,, \\
H &=& -\sin(\beta-\alpha) R + \cos(\beta-\alpha) H^0 \,.
\end{eqnarray}
\end{subequations}
In view of the fact that a $125$ GeV scalar with SM-like properties has already been observed at the LHC, we wish the lighter CP-even mass eigenstate ($h$) to coincide with $H^0$. Then we must require~:
\begin{eqnarray}
\cos(\beta-\alpha) \approx 0 \,.
\end{eqnarray}
In analogy with the 2HDM case \cite{Gunion:2002zf}, this limit can be taken as the {\em decoupling limit} in the context of a 3HDM with an $S_3$ symmetry. We must emphasize though, the term `decoupling limit' does not necessarily imply the heaviness of the additional scalars. Considering Eqs.~(\ref{h0}) and (\ref{H0R}), it is also interesting to note that the state $h^0$, being orthogonal to $H^0$, does not have any trilinear $h^0VV$ ($V=$ $W$,$Z$) coupling. But, in general, it will have flavor changing coupling in the Yukawa sector. This type of neutral massless state with flavor changing fermionic coupling will be ruled out from the well measured values of neutral meson mass differences. This means that the choice $\lambda_4=0$ is phenomenologically unacceptable and we shall not pursue this scenario any further.

\subsection{Case-II ($v_1=\sqrt{3}v_2$)}
This situation has recently been analyzed in \cite{Barradas-Guevara:2014yoa}. We, however, use a convenient parametrization that can provide intuitive insight into the scenario and additionally, we also discuss the possibility of a {\em decoupling limit} in the same way as done in the previous subsection.

The definitions for the angles, $\gamma$ and $\beta$, and the digonalizing  matrix, $X$, remain the same as before. Only difference is that, due to the vev alignment ($v_1=\sqrt{3}v_2$), $\tan\gamma$ ($=\sqrt{3}$) and hence $X$ is determined completely. Now only two of the vevs, $v_2$ and $v_3$ (say), can be considered independent and $\tan\beta$ is given in terms of them as follows~:
\begin{eqnarray}
\tan\beta = \frac{2v_2}{v_3} \,.
\end{eqnarray}
The charged and pseudoscalar mass eigenstates have the same form as before; only the mass eigenvalues get modified due to the presence of $\lambda_4$~:
\begin{subequations}
\begin{eqnarray}
m^2_{1+} &=& -\left\{2\lambda_3\sin^2\beta+ \frac{5}{2}\lambda_4\sin\beta\cos\beta +\frac{1}{2}(\lambda_6+2\lambda_7)\cos^2\beta \right\}v^2 \,, \\
m^2_{2+} &=& -\frac{1}{2}\left\{\lambda_4\tan\beta+(\lambda_6+2\lambda_7) \right\}v^2 \,, \\
m^2_{A1} &=& -\left\{2(\lambda_2+\lambda_3)\sin^2\beta +\frac{5}{2}\lambda_4\sin\beta\cos\beta +2\lambda_7\cos^2\beta \right\}v^2 \,, \\
m^2_{A2} &=& -\left(\frac{1}{2}\lambda_4\tan\beta+2\lambda_7 \right) v^2 \,.
\end{eqnarray}
\end{subequations}
In the presence of $\lambda_4$, analysis of the scalar part will be slightly different~:
\begin{subequations}
\begin{eqnarray}
XM_S^2X^T &=& \begin{pmatrix}
\frac{1}{2}m_{h0}^2 & 0 & 0 \\ 0 & A_S & -B_S \\ 0& -B_S &  C_S
\end{pmatrix} \,, \\
{\rm where,} ~~~ A_S &=& (\lambda_1+\lambda_3)v^2\sin^2\beta +\frac{3}{4}\lambda_4v^2 \sin\beta\cos\beta \,, \\
B_S &=& -\frac{1}{2} \left\{\frac{3}{2}\lambda_4\sin^2\beta + (\lambda_5+\lambda_6+2\lambda_7)\sin\beta\cos\beta \right\}v^2 \,, \\
C_S &=& -\frac{\lambda_4}{4}v^2\sin^2\beta\tan\beta + \lambda_8 v^2\cos^2\beta \,.
\end{eqnarray}
\end{subequations}
The state, $h^0$, will no longer be massless, in fact,
\begin{eqnarray}
m_{h0}^2 = -\frac{9}{2}\lambda_4 v^2\sin\beta\cos\beta \,.
\end{eqnarray}
The angle $\alpha$, which was used to rotate from $(h'_2,~h_3)$ basis to the physical $(H,~h)$ basis, should be redefined as~:
\begin{eqnarray}
 \tan2\alpha &=&\frac{2B_S}{A_S-C_S} \,,
\end{eqnarray}
and corresponding mass eigenvalues should have the following expressions~:
\begin{subequations}
\begin{eqnarray}
m_H^2 &=& (A_S+C_S)+\sqrt{(A_S-C_S)^2+4B_S^{2}} \,, \\
m_h^2 &=& (A_S+C_S)-\sqrt{(A_S-C_S)^2+4B_S^{2}} \,.
\end{eqnarray}
\end{subequations}

The conclusion of the previous subsection that in the decoupling limit, $\cos(\beta-\alpha)=0$, $h$  possesses SM-like gauge and Yukawa couplings, still holds. It should be emphasized that the Yukawa couplings of $h$ in this limit, resembles that of the SM, do not depend on the transformation properties of the fermions under $S_3$. Also, the self couplings of $h$ coincides with the corresponding SM expressions in the decoupling limit~:
\begin{eqnarray}
\mathscr{L}^{\rm self}_h = -\frac{m_h^2}{2v} h^3 -\frac{m_h^2}{8v^2} h^4 \,.
\end{eqnarray}
 Similar to the case described in the previous subsection, $h^0$ will not have any $h^0VV$ ($V=W$, $Z$) couplings, but in the present scenario, we may identify a symmetry which forbids such couplings. Note that, when the specified relation between $v_1$ and $v_2$ is taken, there exists a two dimensional representation of $\mathbb{Z}_2$~:
\begin{eqnarray}
 \begin{pmatrix}1 & 0 \\ 0 & 1 \end{pmatrix} \,,~~
 \frac{1}{2}\begin{pmatrix}1 & \sqrt{3} \\ \sqrt{3} & -1 \end{pmatrix}\,, 
\end{eqnarray}
which was initially a subgroup of the original $S_3$ symmetry, remains intact even after the spontaneous symmetry breaking, {\it i.e.}, the vacuum is invariant under this $\mathbb{Z}_2$ symmetry. This allows us to assign a $\mathbb{Z}_2$ parity for different physical states and this should be conserved in the theory. The state $h^0$ is odd under this $\mathbb{Z}_2$ and this is what forbids it to couple with the $VV$ pair. In fact, using the assignments of Table~\ref{table1}, together with CP symmetry, many of the scalar self couplings can be inferred to be zero.

\begin{table}
\begin{center}
\begin{tabular}{|c|c|}
\hline
Physical States &  Transformation under $\mathbb{Z}_2$ \\
\hline\hline
$h^{0}$, $H_{1}^{\pm}$, $A_{1}$  & Odd \\
\hline
$H^{0}$, $R$, $H_{2}^{\pm}$, $A_{2}$ & Even \\
\hline
\end{tabular}
\end{center}
\caption{\em $\mathbb{Z}_2$ parity assignments to the physical mass eigenstates.}
\label{table1}
\end{table}

In connection with the number of independent parameters in the Higgs potential, we note that there were ten to start with ($\mu_{1,3}$ and $\lambda_{1,2, \dots, 8}$). $\mu_1$ and $\mu_3$ can be traded for $v_2$ and $v_3$ or, equivalently for $v$ and $\tan\beta$. The remaining eight $\lambda$s can be traded for seven physical Higgs masses and $\alpha$. The connections are given below~:
\begin{subequations}
\begin{eqnarray}
\lambda_1 &=& \frac{1}{2v^2\sin^2\beta} \left\{\left(m_h^2\cos^2\alpha+m_H^2\sin^2\alpha\right)+ \left(m_{1+}^2 -m_{2+}^2\cos^2\beta -\frac{1}{9} m_{h0}^2 \right) \right\} \,, \\
\lambda_2 &=& \frac{1}{2v^2\sin^2\beta} \left\{(m_{1+}^2-m_{A1}^2)- (m_{2+}^2-m_{A2}^2)\cos^2\beta \right\} \,, \\
\lambda_3 &=& \frac{1}{2v^2\sin^2\beta} \left(\frac{4}{9} m_{h0}^2 +m_{2+}^2\cos^2\beta - m_{1+}^2\right) \,, \\
\lambda_4 &=& -\frac{2}{9} \frac{m_{h0}^2}{v^2}\frac{1}{\sin\beta\cos\beta} \,, \\
\lambda_5 &=& \frac{1}{v^2} \left\{\frac{\sin\alpha\cos\alpha}{\sin\beta\cos\beta}\left(m_H^2-m_h^2 \right) +2 m_{2+}^2 +\frac{1}{9}\frac{m_{h0}^2}{\cos^2\beta} \right\} \,, \\
\lambda_6 &=& \frac{1}{v^2}\left(\frac{1}{9}\frac{m_{h0}^2}{\cos^2\beta}+m_{A2}^2-2m_{2+}^2 \right) \,, \\
\lambda_7 &=& \frac{1}{2v^2}\left(\frac{1}{9}\frac{m_{h0}^2}{\cos^2\beta}-m_{A2}^2 \right) \,, \\
\lambda_8 &=& \frac{1}{2v^2\cos^2\beta}\left\{\left(m_h^2\sin^2\alpha+m_H^2\cos^2\alpha \right) -\frac{1}{9} m_{h0}^2\tan^2\beta \right\} \,.
\end{eqnarray}
\end{subequations}
In passing, we wish to state that for the analysis purpose we will always be working in the decoupling limit with $v_1=\sqrt{3}v_2$.

\begin{figure} 
%
\includegraphics[scale=0.8]{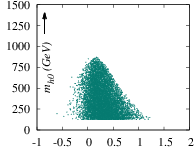} 
\includegraphics[scale=0.8]{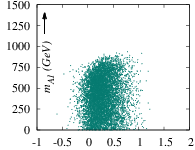}
\includegraphics[scale=0.8]{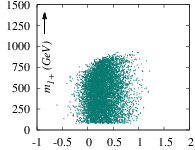}
\centerline{ \null \hfill $\log_{10}(\tan \beta) ~~ \rightarrow$ \quad}

\vspace{4mm}
%
\includegraphics[scale=0.8]{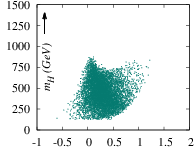} 
\includegraphics[scale=0.8]{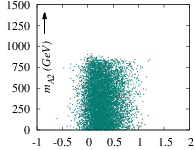}
\includegraphics[scale=0.8]{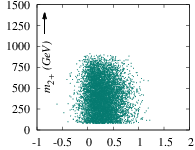}
\centerline{ \null \hfill $\log_{10}(\tan \beta) ~~ \rightarrow$ \quad}

\caption{\em (Case-II) Regions allowed from unitarity and stability. We have fixed $m_h$ at 125 GeV and taken $m_{1+},~m_{2+} > 80$\,GeV and $m_H, m_{h0} > m_h$.}
\label{f:planes}
\end{figure}

\section{Constraints from unitarity} \label{s:unitarity}
In this context, the pioneering work has been done by Lee, Quigg and Thacker (LQT)  \cite{Lee:1977eg}. They have analyzed
several two body scatterings involving longitudinal gauge bosons and physical Higgs in the SM. All such scattering amplitudes are proportional to Higgs quartic coupling in the high energy limit. The $\ell=0$ partial wave amplitude
$(a_0)$ is then extracted from these amplitudes and cast in the form of an S-matrix having different two-body states
as rows and columns. The largest eigenvalue of this matrix is bounded by the unitarity constraint, $|a_0 | < 1$.
This restricts the quartic Higgs self coupling and therefore the Higgs mass to a maximum value.

The procedure has been extended to the case of a 2HDM scalar potential \cite{Maalampi:1991fb,Kanemura:1993hm,Akeroyd:2000wc,Horejsi:2005da}. We take it one step further and apply it in the context of 3HDMs.
Here also same types of two body scattering channels are considered. Thanks to the equivalence theorem \cite{Pal:1994jk,Horejsi:1995jj}, we can use unphysical Higgses instead of actual longitudinal components of the gauge bosons when considering the
high energy limit. So, we can use the Goldstone-Higgs potential of \Eqn{potential} for this analysis. Still it will be a
much involved calculation. But we notice that the diagrams containing trilinear vertices will be suppressed by
a factor of $E^2$ coming from the intermediate propagator. Thus they do not contribute at high energies, − only
the quartic couplings contribute. Clearly the physical Higgs masses that could come from the propagators, do
not enter this analysis. Since we are interested only in the eigenvalues of the S-matrix, this allows us to work
with the original fields of \Eqn{quartic} instead of the physical mass eigenstates. After an inspection of all the neutral and charged two-body channels, we find the following eigenvalues to be bounded from unitarity~:
%
%
\begin{eqnarray}
|a_i^\pm|,~|b_i| \le 16\pi, ~\mbox{for}~i=1,2,\ldots,6\,.
\end{eqnarray}
The expressions for the individual eigenvalues in terms of $\lambda$s are given below~:
\begin{subequations}
\begin{eqnarray}
a_1^\pm &=& \left(\lambda_1-\lambda_2+ \frac{\lambda_5+\lambda_6}{2}\right) \pm \sqrt{\left(\lambda_1-\lambda_2+ \frac{\lambda_5+\lambda_6}{2}\right)^2 -4\left\{(\lambda_1-\lambda_2)\left(\frac{\lambda_5+\lambda_6}{2}\right) -\lambda_4^2 \right\} } \,, \\
a_2^\pm &=& \left(\lambda_1+\lambda_2+2\lambda_3+\lambda_8\right) \pm \sqrt{\left(\lambda_1+\lambda_2+2\lambda_3+\lambda_8\right)^2 -4\left\{\lambda_8(\lambda_1+\lambda_2+2\lambda_3) -2\lambda_7^2 \right\} } \,, \\
a_3^\pm &=& \left(\lambda_1-\lambda_2+2\lambda_3+\lambda_8\right) \pm \sqrt{\left(\lambda_1-\lambda_2+2\lambda_3+\lambda_8\right)^2 -4\left\{\lambda_8(\lambda_1-\lambda_2+2\lambda_3) -\frac{\lambda_6^2}{2} \right\} } \,, \\
a_4^\pm &=& \left(\lambda_1+\lambda_2+\frac{\lambda_5}{2}+\lambda_7\right) \pm  \sqrt{\left(\lambda_1+\lambda_2+\frac{\lambda_5}{2}+\lambda_7\right)^2 -4\left\{(\lambda_1+\lambda_2)\left(\frac{\lambda_5}{2}+\lambda_7\right) -\lambda_4^2 \right\} } \,, \\
a_5^\pm &=& \left(5\lambda_1-\lambda_2+2\lambda_3+3\lambda_8\right) \nonumber \\
&& \pm \sqrt{\left(5\lambda_1-\lambda_2+2\lambda_3+3\lambda_8\right)^2 -4\left\{3\lambda_8(5\lambda_1-\lambda_2+2\lambda_3) -\frac{1}{2}(2\lambda_5+\lambda_6)^2 \right\} } \,, \\
a_6^\pm &=& \left(\lambda_1+\lambda_2+4\lambda_3+\frac{\lambda_5}{2}+\lambda_6+3\lambda_7\right) \nonumber \\
&& \pm \sqrt{\left(\lambda_1+\lambda_2+4\lambda_3+\frac{\lambda_5}{2}+\lambda_6+3\lambda_7\right)^2 -4\left\{(\lambda_1+\lambda_2+4\lambda_3)\left(\frac{\lambda_5}{2}+\lambda_6+3\lambda_7\right) -9\lambda_4^2 \right\} } \,, \\
b_1 &=& \lambda_5+2\lambda_6-6\lambda_7 \,, \\
b_2 &=& \lambda_5 -2\lambda_7 \,, \\
b_3 &=& 2(\lambda_1-5\lambda_2-2\lambda_3) \,, \\
b_4 &=& 2(\lambda_1-\lambda_2-2\lambda_3) \,, \\
b_5 &=& 2(\l_1+\l_2-2\l_3) \,, \\
b_6 &=& \lambda_5-\lambda_6\,.
\end{eqnarray}
\label{unitarity eq}
\end{subequations}
In passing, we remark that the perturbativity criteria, $|\lambda_i| < 4\pi$, coming from the requirement that the leading order contribution to the physical amplitude must have higher magnitude than the subleading order, may have some ambiguity in this context. This is due to the fact the individual $\lambda$s do not appear in the quartic couplings involving the physical scalars. Hence the combination of $\lambda$s, that constitute the physical couplings,  should be used for this purpose and it does not necessarily imply that the individual $\lambda$s should be bounded. We have presented here the exact constraints on $\lambda$s which should be satisfied for unitarity not to be violated.

Eqs.~(\ref{stability}) and (\ref{unitarity eq}) can be used to put limits on the physical Higgs masses. For this purpose, we work in the decoupling limit taking the lightest scalar ($h$) to be the SM-like Higgs that has been found at the LHC and we set its mass at 125 GeV. We also assume the charged scalars ($m_{1+}$ and $m_{2+}$) to be heavier than 80 GeV to respect the direct search bound from LEP2 \cite{Searches:2001ac}. To collect sufficient number of data points we have generated fifty million random sets of \{$\tan\beta,~m_{h0},~m_{H},~m_{A1},~m_{A2},~m_{1+},~m_{2+}$\} by varying $\tan\beta$ from 0.1 to 100 and filter them through the combined constraints from unitarity and stability. The sets that survive the filtering are plotted in Figure~\ref{f:planes}. The bounds that follow from these figures are listed below~:
\begin{itemize}
\item $\tan\beta~\in$  [0.3, 17],
\item $m_{h0}<$ 870 GeV, $m_{H}<$ 880 GeV, $m_{A1}<$ 940 GeV, $m_{A2}<$ 910 GeV, $m_{1+}<$ 940, $m_{2+}<$ 910 GeV.
\end{itemize}
It is interesting to note that, if the observed scalar at the LHC has its root in the S3HDM, then there must be several other nonstandard scalars with masses below 1 TeV.

\section{Impact on loop induced Higgs decays} \label{s:decay}
\begin{figure}
\includegraphics[scale=0.4]{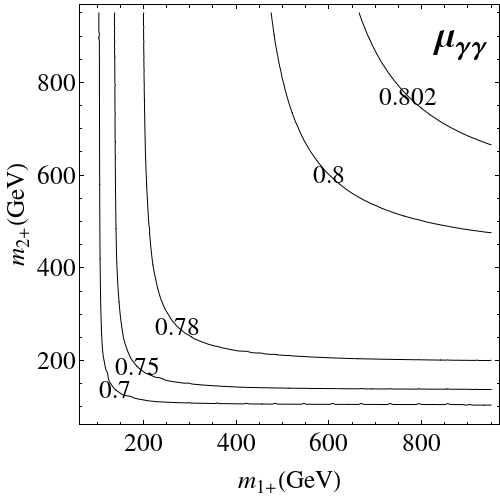}~~~~~~~
\includegraphics[scale=0.4]{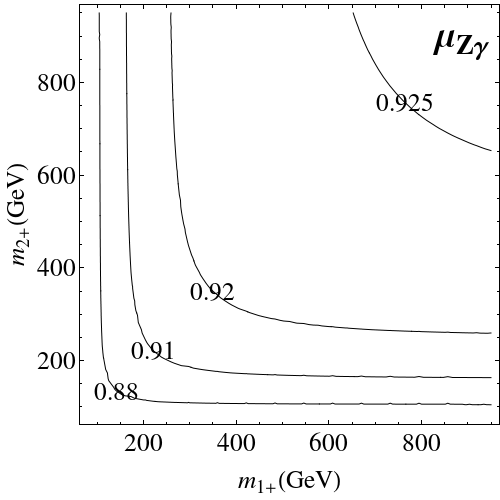}
\caption{\em Signal strengths for diphoton and Z-photon decay modes within the allowed range for charged Higgs masses.}
\label{f:decay}
\end{figure}
As already has been pointed out, in the decoupling limit the lightest scalar ($h$) couples with fermions and gauge bosons exactly in the SM way. Consequently, the production cross section as well as tree level decay branching ratios will not alter from their respective SM values. However, the loop induced decay modes like, $h\to \gamma \gamma$ and $h\to Z \gamma$, will pick up additional contributions due to the presence of nonstandard charged scalar loops. Note that the change in total Higgs decay width will be negligibly small as the branching fractions of such decays are tiny.

To display the contribution of the charged scalar loops to the decay amplitudes in a convenient form, we define dimensionless parameters, $\kappa_{i}~(i=1,2)$, in the following way~:
\begin{eqnarray}
g_{hH_{i}^{+}H_{i}^{-}}=\kappa_{i}\frac{gm_{i+}^{2}}{M_{W}}\,.
\label{defkappa}
\end{eqnarray}
The standard expression for the diphoton decay width is given by \cite{Gunion:1989we}:
\begin{eqnarray}
 \Gamma (h\to \gamma\gamma) = \frac{\alpha^2g^2}{2^{10}\pi^3}
 \frac{m_h^3}{M_W^2} \Big|\mathcal{F}_W + \frac{4}{3}\mathcal{F}_t  + \sum_{i=1}^{2}\kappa_{i} \mathcal{F}_{i+} \Big|^2
 \,, 
\label{h2gg}
\end{eqnarray}
where, using the notation $\tau_x \equiv (2m_x/m_h)^2$, the expressions for $\mathcal{F}_W$, $\mathcal{F}_t$ and $\mathcal{F}_{i+}~(i=1,2)$ are given by,
\begin{subequations}
\begin{eqnarray}
 \mathcal{F}_W &=& 2+3\tau_W+3\tau_W(2-\tau_W)f(\tau_W) \,,  \\
 \mathcal{F}_t &=& -2\tau_t \big[1+(1-\tau_t)f(\tau_t)\big] \,,  \\
 \mathcal{F}_{i+} &=& -\tau_{i+} \big[ 1-\tau_{i+}f(\tau_{i+}) \big]\,.
\end{eqnarray}
\end{subequations}
For the values of masses that we are dealing with, makes $\tau_{x} > 1$ for $x = W,~t,~H_{i}^{\pm}$ and then 
\begin{eqnarray}
f(\tau) =
\left[\sin^{-1}\left(\sqrt{1/\tau}\right)\right]^2 \,.
\label{f}
\end{eqnarray}
The decay width for $h\to Z\gamma$ is given by~:
\begin{eqnarray}
 \Gamma (h\to Z\gamma) = \frac{\alpha^2g^2}{2^{9}\pi^3}
 \frac{m_h^3}{M_W^2} \Big|\mathcal{A}_W + \mathcal{A}_t  + \sum_{i=1}^{2}\kappa_{i} \mathcal{A}_{i+}\Big|^2
 \left(1-\frac{M_Z^2}{m_h^2}\right)^3 \,,
\label{h2Zg}
\end{eqnarray}
where, using $\eta_x = (2m_x/M_Z)^2$, the expressions for $\mathcal{A}_W$, $\mathcal{A}_t$ and $\mathcal{A}_{i+}$ are given by~\cite{Gunion:1989we},
\begin{subequations}
\begin{eqnarray}
 \mathcal{A}_W &=& \cot \theta_w\bigg[ 4(\tan^2\theta_w - 3)I_2(\tau_W,\eta_W)
 \nonumber \\*
&& \null +\bigg\{ \left(5+\frac{2}{\tau_W}\right) -
 \left(1+\frac{2}{\tau_W}\right)\tan^2\theta_w \bigg\}
 I_1(\tau_W,\eta_W)\bigg] \,, 
 \\ 
 \mathcal{A}_t &=&
 \frac{4\Big(\frac{1}{2}-\frac{4}{3}\sin^2\theta_w\Big)}{\sin\theta_w
   \cos\theta_w} \; \Big[I_2(\tau_t,\eta_t)-I_1(\tau_t,\eta_t) \Big] \,, \\
 \mathcal{A}_{i+} &=& \frac{(2\sin^2\theta_w-1)}{\sin\theta_w \cos\theta_w} \;
 I_1(\tau_{i+},\eta_{i+}) \,.
\end{eqnarray}
\end{subequations}
The functions $I_1$ and $I_2$ are defined as,
\begin{subequations}
\begin{eqnarray}
 I_1(\tau,\eta) &=& \frac{\tau \eta}{2(\tau -\eta)} +
 \frac{\tau ^2\eta^2}{2(\tau -\eta)^2}\Big[f(\tau )-f(\eta)\Big]
 +\frac{\tau ^2\eta}{(\tau -\eta)^2}\Big[g(\tau )-g(\eta)\Big]  \,, \\
 I_2(\tau ,\eta) &=& -\frac{\tau \eta}{2(\tau -\eta)}\Big[f(\tau
   )-f(\eta)\Big] \,,  
\end{eqnarray}
\end{subequations}
where the function $f$ has the same definition as in \Eqn{f}. Since $\tau_x,~\eta_x > 1$ for $x = W,~t,~H_{i}^{\pm}$, the function $g$ takes the
following form:
\begin{eqnarray}
 g(x) = \sqrt{x-1}\sin^{-1}\left(\sqrt{1/x}\right) \,.
\end{eqnarray}
In the decoupling limit, the parameters $\kappa_{i}~(i=1,2)$, which appear in
Eqs.\ (\ref{defkappa}), (\ref{h2gg}) and (\ref{h2Zg}) are given by,
\begin{eqnarray}
\kappa_{i} = -\left(1 + \frac{m_h^2}{2m_{i+}^2}\right)\,.
\label{kappa}
\end{eqnarray}
In our case, the signal strengths $\mu_{\gamma\gamma}$ and $\mu_{Z\gamma}$,
defined through the equations,
\begin{eqnarray}
 \mu_{\gamma\gamma} &=& {\sigma(pp\to h) \over \sigma^{\rm
     SM}(pp\to h)} \cdot {\mbox{BR} (h \to 
   \gamma\gamma) \over \mbox{BR}^{\rm SM} (h \to \gamma\gamma)} \,,
 \\ 
 \mu_{Z\gamma} &=& {\sigma(pp\to h) \over \sigma^{\rm
     SM}(pp\to h)} \cdot {\mbox{BR} (h \to 
   Z\gamma) \over \mbox{BR}^{\rm SM} (h \to Z\gamma)} \,,
\label{mu}
\end{eqnarray}
assume the following forms:
\begin{eqnarray}
 \mu_{\gamma\gamma} &=&  \frac{\Gamma(h \to \gamma\gamma)}{\Gamma^{\rm
     SM}(h\to \gamma\gamma)} 
 = \frac{\Big|\mathcal{F}_W + \frac{4}{3}\mathcal{F}_t  +
    \sum_{i=1}^{2}\kappa_{i} \mathcal{F}_{i+}\Big|^2}{\Big|\mathcal{F}_W + \frac{4}{3}\mathcal{F}_t \Big|^2} \,,
 \\ 
 \mu_{Z\gamma} &=&  \frac{\Gamma(h \to
  Z\gamma)}{\Gamma^{\rm SM}(h\to Z\gamma)} = \frac{\Big|\mathcal{A}_W + \mathcal{A}_t  +
   \sum_{i=1}^{2}\kappa_{i} \mathcal{A}_{i+} \Big|^2}{ \Big|\mathcal{A}_W + \mathcal{A}_t \Big|^2} \,.
\end{eqnarray}
As the charged Higgs becomes heavy, the quantity $\mathcal{F}_{i+}$, for example, saturates to $\frac{1}{3}$.  So the decoupling of charged Higgs from loop induced Higgs decay depends on how $\kappa_{i}$ behaves with increasing $m_{i+}$. It follows from \Eqn{kappa} that $\kappa_{i}\to -1$ if $m_{i+}\gg m_{h}$. Consequently, the charged Higgs never decouples from the diphoton or $Z$-photon decay amplitudes. In fact, it reduces the decay widths from their corresponding SM expectations. These features have been displayed in Figure~\ref{f:decay} where we have made a contour plot by varying the charged Higgs masses within the allowed ranges coming from unitarity and vacuum stability. We find that $\mu_{\gamma \gamma}$ and $\mu_{Z \gamma}$ should lie within  [0.42, 0.80] and [0.73, 0.93] for $m_{1+}\in [80, 950]$ and $m_{2+}\in [80, 950]$. We must admit though, this nondecoupling of charged scalar is not a unique feature of a S3HDM as it is also known to be present in the context of a 2HDMs \cite{Djouadi:1996yq, Arhrib:2003ph, Bhattacharyya:2013rya, Ferreira:2014naa}. Currently the ATLAS data favor an enhancement whereas the data from CMS favor a suppression in the diphoton decay channel \cite{pdg}. Thus a precise measurement of the diphoton and $Z$-photon signal strengths can pin down the difference between the SM Higgs and a SM-like Higgs arising from an extended scalar sector. 

\section{Conclusions} \label{s:concl}
In this article we have analyzed in detail the scalar sector of an S3HDM. Our findings are listed below~:
\begin{itemize}
\item The minimization of the scalar potential leads to a specific relation between the vevs of the first two doublets, $v_{1}=\sqrt{3} v_{2}$ in particular.

\item  In this limit we find a $\mathbb{Z}_{2}$ subgroup of $S_{3}$ that remains unbroken even after the spontaneous symmetry breaking. The different scalar mass eigenstates can then be assigned with appropriate $\mathbb{Z}_{2}$ parity which can help us understand why certain couplings do not appear in the theory.

\item Additionally, we have identified a decoupling limit for this model where the lightest CP-even scalar has the exact same coupling as the SM Higgs with the other SM particles.

\item We have also derived the exact tree unitarity constraints and exploited them, in the decoupling limit, to put new bounds on the physical nonstandard Higgs masses, which we consider to be an important development in the multi-Higgs context.

\item From unitarity and stability $\tan \beta$ is likely to be in the range [0.3,17] and all the nonstandard Higgs masses lie below 1 TeV.

\item Regarding the decay of the SM-like $S_{3}$ Higgs, we have observed that the charged Higgs never decouples from the diphoton or $Z$-photon decay modes. The additional contributions from the charged Higgs loops to the decay amplitudes actually reduces the signal strengths of these modes. Although this depletion may not be a unique property of this scenario, but any statistically significant enhancement in $h\to \gamma \gamma$ and $h\to Z \gamma$ modes will certainly disfavor the possibility of an SM-like Higgs arising from an S3HDM.
\end{itemize}


\paragraph*{Acknowledgements\,:} We thank Gautam Bhattacharyya, Palash B. Pal and Amitava Raychaudhuri for their insightful comments at different stages of this work. Technical help from Arindam Chatterjee is also gratefully acknowledged.  UKD acknowledges the funding available from the Department of Atomic Energy, Government of India, for the Regional Centre for Accelerator-based Particle Physics, Harish-Chandra Research Institute. DD thanks the Department of Atomic Energy, India, for financial support.

\appendix

\section{Finding the unitarity constraints} \label{s:appen}
In this appendix we present a detailed account of our discussions regarding unitarity bounds  in Section~\ref{s:unitarity}. Any scattering amplitude can be expanded in terms of the partial waves as follows~:
\begin{equation}
\mathcal{M}(\theta) = 16 \pi\sum\limits_{\ell=0}^{\infty} a_\ell (2\ell+1)P_\ell (\cos\theta) \,,
\label{def:feynman}
\end{equation}
where, $\theta$ is the scattering angle and $P_\ell (x)$ is the Legendre polynomial of order $\ell$. The prescription is simple~: once we calculate the Feynman amplitude of a certain $2\to 2$ scattering process, each of the partial wave amplitude ($a_\ell$), in \Eqn{def:feynman}, can be extracted by using the orthonormality of the Legendre polynomials. As argued in Section~\ref{s:unitarity}, only the dimensionless quartic couplings will contribute to the amplitudes under consideration at high energies. For this, only $\ell=0$ partial amplitude ($a_0$) will receive nonzero contribution from the leading order term in the scattering amplitude. It is our purpose, then, to find the expressions of $a_0$ for every possible $2\to 2$ scattering process and cast them in the form of an S-matrix which is constructed by taking the different two-body channels as rows and columns. Unitarity will restrict the magnitude of each of the eigenvalues of this S-matrix to lie below unity. The resultant constraints have been in quoted in \Eqn{unitarity eq}.

First important part of the calculation is to identify all the possible two-particle channels. These two-particle states are made of the fields $w_{k}^{\pm},~h_{k}$ and $z_{k}$ corresponding to the parametrization of \Eqn{e:doub}. For our calculation, we consider neutral two-particle states (e.g., $w_{i}^{+}w_{j}^{-},~h_i h_j,~z_i z_j,~h_i z_j$) and singly charged two-particle states (e.g., $w_{i}^{+}h_j,~w_{i}^{+}z_j$). In general, if we have $n$-number of doublets $\phi_{k}~(k=1,\ldots,n)$ there will be $(3n^2+n)$-number of neutral  and $2n^2$-number of charged two-particle states. Clearly, the dimensions of S-matries formed out of these two-particle states will be a $(3n^2+n)\times(3n^2+n)$  and $2n^2\times 2n^2$ for the neutral and charged cases respectively. The eigenvalues of these matrices should be bounded by the unitarity constraint.

\subsection{Neutral Channels:}
In our case of three Higgs doublets there will be, $3\cdot(3)^2+3=30$ neutral two-particle states and thus the neutral channel S-matrix will be a $30\times 30$ matrix.  The symmetries present in our potential, \Eqn{potential} and a few tricks allows us to get analytical expressions for the eigenvalues of this matrix. The basis of neutral two-particle states (NTPS) are,
\begin{eqnarray*}
&&\{w_{1}^{+}w_{2}^{-},~w_{1}^{+}w_{3}^{-},~w_{2}^{+}w_{1}^{-},~w_{3}^{+}w_{1}^{-},~h_{1}h_{2},~h_{1}h_{3},~z_{1}z_{2},~z_{1}z_{3},~h_{1}z_{2},~h_{1}z_{3},~z_{1}h_{2},~z_{1}h_{3}\}~\rm{and}\\
&&\{h_{3}z_{3},~h_{1}z_{1},~h_{2}z_{2},~h_{3}z_{2},~h_{2}z_{3},~w_{3}^{+}w_{2}^{-},~w_{2}^{+}w_{3}^{-},~h_{2}h_{3},~z_{2}z_{3},~w_{1}^{+}w_{1}^{-},~w_{2}^{+}w_{2}^{-},~w_{3}^{+}w_{3}^{-},\\
&&~~~~~~~~~~\frac{h_{1}h_{1}}{\sqrt{2}},~\frac{h_{2}h_{2}}{\sqrt{2}},~\frac{h_{3}h_{3}}{\sqrt{2}},~\frac{z_{1}z_{1}}{\sqrt{2}},~\frac{z_{2}z_{2}}{\sqrt{2}},~\frac{z_{3}z_{3}}{\sqrt{2}}\}
\end{eqnarray*}
Note that, the states containing two identical bosons contain an additional factor of $\frac{1}{\sqrt{2}}$ due to bose symmetry. We divide the NTPS in two classes. This classification helps us to reduce, as a first level of simplification, the $30\times 30$ matrix to a $12\times 12$ and $18\times 18$ block diagonal form. If a bigger matrix can be block diagonalized in smaller matrices then the calculation of the eigenvalues of the original matrix becomes easier. In the present case this type of block diagonalization is possible due to the very structure of the potential. From the potential, \Eqn{potential} it is evident that transition from two-particle states containing even number of the index `1' into two-particle states having odd number of `1' and vice versa,  are not allowed. This explains why the first set of NTPS above, are completely disentangled from the second set. The $12\times 12$ matrix constructed using the first set of NTPS is given by,
\begin{equation}
\mathcal{M}^{(1)}_{NC}= \begin{pmatrix}
\mathcal{A}_{6\times 6} & \mathcal{B}_{6\times 6} \\ \mathcal{B}^{\dagger}_{6\times 6} & \mathcal{C}_{6\times 6}
\end{pmatrix} \,,
\label{mat}
\end{equation} 
where $\mathcal{A}$, $\mathcal{B}$ and $\mathcal{C}$ are given by,
\begin{eqnarray*}
\mathcal{A} &=&
  \bordermatrix{
  &\m w_1^+ w_2^-&\m  w_1^+ w_3^-&\m w_2^+ w_1^-&\m w_3^+ w_1^-&\m h_1 h_2&\m h_1 h_3 \cr\vbox{\hrule}
\m w_1^+ w_2^- & 2(\l_1-\l_2) & 2\l_4  & 4(\l_2+\l_3) & 2\l_4 & 2\l_3
    & \l_4                     
   \cr
\m w_1^+ w_3^-& 2\l_4 & \l_5+\l_6 & 2\l_4 & 4\l_7 & \l_4 & \frac{\l_6}{2}+\l_7
   \cr
\m w_2^+ w_1^- & 4(\l_2+\l_3) & 2\l_4 & 2(\l_1-\l_2) & 2\l_4 & 2\l_3 & \l_4 
   \cr
\m w_3^+ w_1^- & 2\l_4 & 4\l_7 & 2\l_4 & \l_5+\l_6 & \l_4 & \frac{\l_6}{2}+\l_7  
   \cr 
\m h_1 h_2 & 2\l_3 & \l_4 & 2\l_3 & \l_4 & 2(\l_1 +\l_3) & 3\l_4 
   \cr
\m h_1 h_3 & \l_4 & \frac{\l_6}{2}+\l_7 & \l_4 & \frac{\l_6}{2}+\l_7 & 3\l_4 & \l_5 + \l_6 +2\l_7  \cr
   }\,, \\
\mathcal{B} &=& 
  \bordermatrix{
  &\m z_1 z_2 &\m z_1 z_3 &\m h_1 z_2 &\m h_1 z_3 &\m z_1 h_2 &\m z_1 h_3 \cr\vbox{\hrule}
\m w_1^+ w_2^- & 2\l_3 & \l_4  & -2 i \l_2 & 0 & 2 i \l_2
    & 0                     
   \cr
\m w_1^+ w_3^-& \l_4 & \frac{\l_6}{2}+\l_7 & 0 & \frac{i}{2}(\l_6-2\l_7) & 0 & -\frac{i}{2}(\l_6-2\l_7)
   \cr
\m w_2^+ w_1^- & 2\l_3 & \l_4 & 2 i \l_2 & 0 & -2 i \l_2 & 0 
   \cr
\m w_3^+ w_1^- & \l_4 & \frac{\l_6}{2}+\l_7 & 0 & -\frac{i}{2}(\l_6-2\l_7) & 0 & \frac{i}{2}(\l_6-2\l_7)  
   \cr 
\m h_1 h_2 & 2(\l_2+\l_3) & \l_4 & 0 & 0 & 0 & 0 
   \cr
\m h_1 h_3 & \l_4 & 2\l_7 & 0 & 0 & 0 & 0  \cr
   }\,, \\
\mathcal{B}^{\dagger} &=& 
  \bordermatrix{
  &\m w_1^+ w_2^-&\m  w_1^+ w_3^-&\m w_2^+ w_1^-&\m w_3^+ w_1^-&\m h_1 h_2&\m h_1 h_3 \cr\vbox{\hrule}
\m z_1 z_2 & 2\l_3 & \l_4  & 2 \l_3 & \l_4 & 2 (\l_2+\l_3) & \l_4                     
   \cr
\m z_1 z_3 & \l_4 & \frac{\l_6}{2}+\l_7 & \l_4 & \frac{\l_6}{2}+\l_7 & \l_4 & 2\l_7
   \cr
\m h_1 z_2 & 2i\l_2 & 0 & -2 i \l_2 & 0 & 0 & 0 
   \cr
\m h_1 z_3 & 0 & -\frac{i}{2}(\l_6-2\l_7) & 0 & \frac{i}{2}(\l_6-2\l_7) & 0 & 0  
   \cr 
\m z_1 h_2 & -2 i \l_2 & 0 & 2 i \l_2 & 0 & 0 & 0 
   \cr
\m z_1 h_3 & 0 & \frac{i}{2}(\l_6-2\l_7) & 0 & -\frac{i}{2}(\l_6-2\l_7) & 0 & 0  \cr
   }\,, \\   
   \mbox{and,}\\
 \mathcal{C} &=& 
  \bordermatrix{
  &\m z_1 z_2 &\m z_1 z_3 &\m h_1 z_2 &\m h_1 z_3 &\m z_1 h_2 &\m z_1 h_3 \cr\vbox{\hrule}
\m z_1 z_2 & 2(\l_1+\l_3) & 3\l_4  & 0 & 0 & 0  & 0                    
   \cr
\m z_1 z_3& 3\l_4 & {\scriptstyle \l_5 + \l_6 +2\l_7} & 0 & 0 & 0 & 0
   \cr
\m h_1 z_2 & 0 & 0 & {\scriptstyle 2(\l_1-2\l_2-\l_3)} & \l_4 & 2(\l_2+\l_3) & \l_4 
   \cr
\m h_1 z_3 & 0 & 0 & \l_4 & {\scriptstyle \l_5 + \l_6 -2\l_7} & \l_4 & 2\l_7  
   \cr 
\m z_1 h_2 & 0 & 0 & 2(\l_2+\l_3) & \l_4 & {\scriptstyle 2(\l_1-2\l_2-\l_3)} & \l_4 
   \cr
\m z_1 h_3 & 0 & 0 & \l_4 & 2\l_7 & \l_4 & {\scriptstyle \l_5+\l_6-2\l_7}  \cr
   } \,.
 \end{eqnarray*}   
One can get analytical expressions of the eigenvalues of  $\mathcal{M}^{(1)}_{NC}$ using MATHEMATICA. With reference to \Eqn{unitarity eq}, these are $b_i~(i=1,\ldots,4)$, $a^{\pm}_i~(i=1,4,6)$ with $a_1^\pm$ being twofold degenerate.

Now, the $18\times 18$ matrix constructed using the second set of NTPS is $\mathcal{M}^{(2)}_{NC}$. To decompose it further into block diagonal form, we make use of the CP symmetry. Note that, $w_{2}^{+}w_{3}^{-}$ and $w_{2}^{-}w_{3}^{+}$ do not possess any definite CP properties but the linear combinations of them 
\begin{eqnarray}
w_{23}^-&=&\frac{1}{\sqrt{2}}(-w_{2}^{+}w_{3}^{-}+w_{3}^{+}w_{2}^{-}) \,, ~~{\rm and} \\
w_{23}^+&=&\frac{1}{\sqrt{2}}(w_{2}^{+}w_{3}^{-}+w_{3}^{+}w_{2}^{-}) \,,
\label{CP}
\end{eqnarray}
are CP-odd and CP-even states respectively.  A closer look at the second set of NTPS reveals that the first five states are CP-odd whereas the last eleven states are CP-even. Clearly, if CP is conserved in the Higgs potential, then we may rotate the sixth and seventh states into $w_{23}^-$ and $w_{23}^+$ to assure the block diagonalization. Evidently the matrix, $U$, needed to perform unitary transformation on the original $18\times 18$ matrix, can be constructed as follows~:
\begin{equation}
U = \mbox{Block-diag}[X,Y,Z] \,,
\end{equation}
where,
\begin{equation}
 X = \mathbf{1}_{5\times 5},~~~Y = \frac{1}{\sqrt{2}}\begin{pmatrix}
 1 & -1 \\ 1 & 1 \end{pmatrix},~~~Z = \mathbf{1}_{11\times 11}\,.
\end{equation}
After the unitary transformation, we obtain the new matrix in the block diagonal form as given below,
\begin{equation}
\tilde{\mathcal{M}}^{(2)}_{NC} = U \mathcal{M}^{(2)}_{NC} U^{\dagger} = \begin{pmatrix}
\mathcal{D}_{6\times 6} & \mathbf{0}_{6\times 12} \\ \mathbf{0}_{12\times 6} & \mathcal{E}_{12\times 12}
\end{pmatrix} \,,
\end{equation}
where,
\begin{equation} 
\mathcal{D} = 
  \bordermatrix{ 
  &\m h_3 z_3 &\m h_1 z_1 &\m h_2 z_2 &\m h_3 z_2 &\m h_2 z_3 &\m w_{23}^{-} \cr\vbox{\hrule}
\m h_3 z_3 & 2\l_8 & 2\l_7  & 2\l_7 & 0 & 0 & 0                     
   \cr
\m h_1 z_1& 2\l_7 & 2(\l_1+\l_3) & 2(\l_2+\l_3) & \l_4 & \l_4 & 0
   \cr
\m h_2 z_2 & 2\l_7 & 2(\l_2+\l_3) & 2(\l_1+\l_3) & -\l_4 & -\l_4 & 0 
   \cr
\m h_3 z_2 & 0 & \l_4 & -\l_4 & \l_5 + \l_6 -2\l_7 & 2\l_7 & \frac{-i}{2}(\l_6-2\l_7)  
   \cr 
\m h_2 z_3 & 0 & \l_4 & -\l_4 & 2\l_7 & \l_5 + \l_6 -2\l_7 & \frac{i}{2}(\l_6-2\l_7) 
   \cr
\m w_{23}^{-} & 0 & 0 & 0 & \frac{i}{2}(\l_6-2\l_7) & \frac{-i}{2}(\l_6-2\l_7) & \l_5+\l_6-4\l_7  \cr
   }\,,
\end{equation}
 contains the CP-odd states and has eigenvalues $a^{\pm}_{i},~b_{i}~\mbox{for}~i = 1,2$ which are listed in \Eqn{unitarity eq}. The matrix $\mathcal{H}$ can be written as,
\begin{equation}
\mathcal{E}= \begin{pmatrix}
\mathcal{F}_{6\times 6} & \mathcal{G}^{T}_{6\times 6} \\ \mathcal{G}_{6\times 6} & \mathcal{H}_{6\times 6}
\end{pmatrix} \,,
\end{equation}
where $\mathcal{F}$, $\mathcal{G}$ and $\mathcal{H}$ are given by,
\begin{eqnarray*}
\mathcal{F} &=&
  \bordermatrix{
  &\m w_{23}^{+} &\m  h_2 h_3 &\m  z_2 z_3&\m w_1^+ w_1^-&\m w_2^+ w_2^-&\m w_3^+ w_3^- \cr\vbox{\hrule}
\m w_{23}^{+} & \l_5 + \l_6 +4\l_7 & \frac{\l_6+2\l_7}{\sqrt{2}}  & \frac{\l_6+2\l_7}{\sqrt{2}} & 2\sqrt{2}\l_4 & -2\sqrt{2}\l_4
    & 0                     
   \cr
\m h_2 h_3 & \frac{\l_6+2\l_7}{\sqrt{2}} & \l_5 + \l_6 +2\l_7 & 2\l_7 & \l_4 & -\l_4 & 0
   \cr
\m z_2 z_3 & \frac{\l_6+2\l_7}{\sqrt{2}} & 2\l_7 & \l_5 + \l_6 +2\l_7 & \l_4 & -\l_4 & 0 
   \cr
\m w_1^+ w_1^- & 2\sqrt{2}\l_4 & \l_4 & \l_4 & 4(\l_1+\l_3) & 2(\l_1-\l_2) & \l_5+\l_6  
   \cr 
\m w_2^+ w_2^- & -2\sqrt{2}\l_4 & -\l_4 & -\l_4 & 2(\l_1-\l_2) & 4(\l_1+\l_3) & \l_5+\l_6 
   \cr
\m w_3^+ w_3^- & 0 & 0 & 0 & \l_5+\l_6 & \l_5+\l_6 & 4\l_8  \cr
   }, \\
\mathcal{G}^{T} &=& 
  \bordermatrix{
  &\m \frac{h_1 h_1}{\sqrt{2}} &\m \frac{h_2 h_2}{\sqrt{2}} &\m \frac{h_3 h_3}{\sqrt{2}} &\m \frac{z_1 z_1}{\sqrt{2}} &\m \frac{z_2 z_2}{\sqrt{2}} &\m \frac{z_3 z_3}{\sqrt{2}} \cr\vbox{\hrule}
\m w_{23}^{+} & \l_4 & -\l_4  & 0 & \l_4 & -\l_4 & 0 
   \cr
\m h_2 h_3 & \frac{3\l_4}{\sqrt{2}} & \frac{-3\l_4}{\sqrt{2}} & 0 & \frac{\l_4}{\sqrt{2}} & \frac{-\l_4}{\sqrt{2}} & 0 
   \cr
\m z_2 z_3 & \frac{\l_4}{\sqrt{2}} & -\frac{\l_4}{\sqrt{2}} & 0 & \frac{3\l_4}{\sqrt{2}} & -\frac{3\l_4}{\sqrt{2}} & 0 
   \cr
\m w_1^+ w_1^- & \sqrt{2}(\l_1+\l_3) & \sqrt{2}(\l_1-\l_3) & \frac{\l_5}{\sqrt{2}} & \sqrt{2}(\l_1+\l_3) & \sqrt{2}(\l_1-\l_3) & \frac{\l_5}{\sqrt{2}} 
   \cr 
\m w_2^+ w_2^- & \sqrt{2}(\l_1-\l_3) & \sqrt{2}(\l_1+\l_3) & \frac{\l_5}{\sqrt{2}} & \sqrt{2}(\l_1-\l_3) & \sqrt{2}(\l_1+\l_3) & \frac{\l_5}{\sqrt{2}} 
   \cr
\m w_3^+ w_3^- & \frac{\l_5}{\sqrt{2}} & \frac{\l_5}{\sqrt{2}} & \sqrt{2}\l_8 & \frac{\l_5}{\sqrt{2}} & \frac{\l_5}{\sqrt{2}} & \sqrt{2}\l_8  \cr
   }\,, \\   
\mathcal{G} &=& 
  \bordermatrix{
  &\m w_{23}^{+} &\m  h_2 h_3 &\m  z_2 z_3&\m w_1^+ w_1^-&\m w_2^+ w_2^-&\m w_3^+ w_3^- \cr\vbox{\hrule}
\m \frac{h_1 h_1}{\sqrt{2}} & \l_4 & \frac{3\l_4}{\sqrt{2}}  & \frac{\l_4}{\sqrt{2}} & \sqrt{2}(\l_1+\l_3) & \sqrt{2}(\l_1-\l_3) & \frac{\l_5}{\sqrt{2}}                     
   \cr
\m \frac{h_2 h_2}{\sqrt{2}} & -\l_4 & \frac{-3\l_4}{\sqrt{2}}  & \frac{-\l_4}{\sqrt{2}} & \sqrt{2}(\l_1-\l_3) & \sqrt{2}(\l_1+\l_3) & \frac{\l_5}{\sqrt{2}} 
   \cr
\m \frac{h_3 h_3}{\sqrt{2}} & 0 & 0 & 0 & \frac{\l_5}{\sqrt{2}} & \frac{\l_5}{\sqrt{2}} & \sqrt{2}\l_8 
   \cr
\m \frac{z_1 z_1}{\sqrt{2}} & \l_4 & \frac{\l_4}{\sqrt{2}}  & \frac{3\l_4}{\sqrt{2}} & \sqrt{2}(\l_1+\l_3) & \sqrt{2}(\l_1-\l_3) & \frac{\l_5}{\sqrt{2}} 
   \cr 
\m \frac{z_2 z_2}{\sqrt{2}} & -\l_4 & \frac{-\l_4}{\sqrt{2}}  & \frac{-3\l_4}{\sqrt{2}} & \sqrt{2}(\l_1-\l_3) & \sqrt{2}(\l_1+\l_3) & \frac{\l_5}{\sqrt{2}} 
   \cr
\m \frac{z_3 z_3}{\sqrt{2}} & 0 & 0 & 0 & \frac{\l_5}{\sqrt{2}} & \frac{\l_5}{\sqrt{2}} & \sqrt{2}\l_8  \cr
   }\,, ~~~{\rm and} \\
 \\
\mathcal{H} &=& 
  \bordermatrix{
  &\m \frac{h_1 h_1}{\sqrt{2}} &\m \frac{h_2 h_2}{\sqrt{2}} &\m \frac{h_3 h_3}{\sqrt{2}} &\m \frac{z_1 z_1}{\sqrt{2}} &\m \frac{z_2 z_2}{\sqrt{2}} &\m \frac{z_3 z_3}{\sqrt{2}} \cr\vbox{\hrule}
\m \frac{h_1 h_1}{\sqrt{2}} & 3(\l_1+\l_3) & \l_1+\l_3 & \frac{\l_5+\l_6+2\l_7}{2} & \l_1+\l_3 & \l_1-2\l_2-\l_3 & \frac{\l_5+\l_6-2\l_7}{2}
   \cr
\m \frac{h_2 h_2}{\sqrt{2}} & \l_1+\l_3 & 3(\l_1+\l_3) & \frac{\l_5+\l_6+2\l_7}{2} & \l_1-2\l_2-\l_3 & \l_1+\l_3 & \frac{\l_5+\l_6-2\l_7}{2} 
   \cr
\m \frac{h_3 h_3}{\sqrt{2}} & \frac{\l_5+\l_6+2\l_7}{2} & \frac{\l_5+\l_6+2\l_7}{2} & 3\l_8 & \frac{\l_5+\l_6-2\l_7}{2} & \frac{\l_5+\l_6-2\l_7}{2} & \l_8 
   \cr
\m \frac{z_1 z_1}{\sqrt{2}} & \l_1+\l_3 & \l_1-2\l_2-\l_3  & \frac{\l_5+\l_6-2\l_7}{2} & 3(\l_1+\l_3) & \l_1+\l_3 & \frac{\l_5+\l_6+2\l_7}{2} 
   \cr 
\m \frac{z_2 z_2}{\sqrt{2}} & \l_1-2\l_2-\l_3 & \l_1+\l_3  & \frac{\l_5+\l_6-2\l_7}{2} & \l_1+\l_3 & 3(\l_1+\l_3) & \frac{\l_5+\l_6+2\l_7}{2} 
   \cr
\m \frac{z_3 z_3}{\sqrt{2}} & \frac{\l_5+\l_6-2\l_7}{2} & \frac{\l_5+\l_6-2\l_7}{2} & \l_8 & \frac{\l_5+\l_6+2\l_7}{2} & \frac{\l_5+\l_6+2\l_7}{2} & 3\l_8  \cr
   }\,.    
\end{eqnarray*}   
The eigenvalues of $\mathcal{E}$ can be found to be $a^{\pm}_{i}~(i=1,\ldots,6)$ which are listed in \Eqn{unitarity eq}. Thus by obtaining the eigenvalues of $\mathcal{D}$ and $\mathcal{E}$ we get all the eighteen eigenvalues of $\mathcal{M}^{(2)}_{NC}$. Earlier we obtained the twelve eigenvalues of $\mathcal{M}^{(1)}_{NC}$. So we get all thirty eigenvalues of the $30\times 30$ neutral channel S-matrix. 

\subsection{Charged Channels:}
There will be $2\cdot(3)^2=18$ charged two-particle states (CTPS) in the case of three Higgs doublets. That is why the charged channel S-matrix will be an $18\times 18$ matrix. We write the basis of CTPS as,
\begin{eqnarray*}
&&\{ w_{1}^{+} h_2,~w_{1}^{+} h_3,~w_{1}^{+} z_2,~w_{1}^{+} z_3,~w_{2}^{+} h_1,~w_{2}^{+} z_1,~w_{3}^{+} h_1,~w_{3}^{+} z_1,~w_{1}^{+} h_1,~w_{1}^{+} z_1,~w_{2}^{+} h_2,~w_{2}^{+} h_3,\\
&&~~~~~~~~~~w_{2}^{+} z_2,~w_{2}^{+} z_3,~w_{3}^{+} h_2,~w_{3}^{+} h_3,~w_{3}^{+} z_2~w_{3}^{+} z_3\}\,.
\end{eqnarray*}
For reasons explained in the text before \Eqn{mat}, this choice of basis will lead to a $(8\times 8)\oplus (10\times 10) $ block-diagonal S-matrix in the charged sector as follows~:
\begin{equation}
\mathcal{M}_{CC}= \begin{pmatrix}
\mathcal{J}_{8\times 8} & \mathbf{0}_{8\times 10} \\ \mathbf{0}_{10\times 8} & \mathcal{K}_{10\times 10}
\end{pmatrix} \,.
\end{equation}
Clearly if we can find the eigenvalues of the matrices $\mathcal{J}$ and $\mathcal{K}$ we get all the eigenvalues of $\mathcal{M}_{CC}$.
The matrix $\mathcal{J}$ is given by,
\begin{eqnarray*}
\mathcal{J} = 
  \bordermatrix{
  &\m w_{1}^{+} h_2 &\m  w_{1}^{+} h_3 &\m w_{1}^{+} z_2 &\m w_{1}^{+} z_3 &\m w_{2}^{+} h_1 &\m w_{2}^{+} z_1 &\m w_{3}^{+} h_1 &\m w_{3}^{+} z_1 \cr\vbox{\hrule}
\m w_{1}^{+} h_2 & 2(\l_1 - \l_3) & \l_4  & 0 & 0 & 2\l_3 & 2i\l_2 & \l_4 & 0                     
   \cr
\m w_{1}^{+} h_3 & \l_4 & \l_5 & 0 & 0 & \l_4 & 0 & \frac{\l_6+2\l_7}{2} & \frac{-i(\l_6-2\l_7)}{2} 
   \cr
\m w_{1}^{+} z_2 & 0 & 0 & 2(\l_1 - \l_3) & \l_4 & -2i\l_2 & 2\l_3 & 0 & \l_4
   \cr
\m w_{1}^{+} z_3 & 0 & 0 & \l_4 & \l_5 & 0 & \l_4 & \frac{i(\l_6-2\l_7)}{2} & \frac{\l_6+2\l_7}{2} 
   \cr 
\m w_{2}^{+} h_1 & 2\l_3 & \l_4 & 2i\l_2 & 0 & 2(\l_1-\l_3) & 0 & \l_4 & 0 
   \cr
\m w_{2}^{+} z_1 & -2i\l_2 & 0 & 2\l_3 & \l_4 & 0 & 2(\l_1-\l_3) & 0 & \l_4  \cr
\m w_{3}^{+} h_1 & \l_4 & \frac{\l_6+2\l_7}{2} & 0 & \frac{-i(\l_6-2\l_7)}{2} & \l_4 & 0 & \l_5 & 0 \cr
\m w_{3}^{+} z_1 & 0 & \frac{i(\l_6-2\l_7)}{2} & \l_4 & \frac{\l_6+2\l_7}{2} & 0 & \l_4 & 0 & \l_5 \cr
   }\,. \\ 
\end{eqnarray*}
The eigenvalues of this matrix are $a_{i}^{\pm}~(i=1,4)$ and $b_{i}~(i=2,4,5,6)$ which are listed in \Eqn{unitarity eq}. The matrix $\mathcal{K}$ can be written as,
\begin{equation}
\mathcal{K}= \begin{pmatrix}
\mathcal{P}_{5\times 5} & \mathcal{Q}_{5\times 5} \\ \mathcal{Q}_{5\times 5}^{\dagger} & \mathcal{R}_{5\times 5}
\end{pmatrix} \,,
\end{equation}
where $\mathcal{P}, ~\mathcal{Q}$ and $\mathcal{R}$ are given by,
\begin{eqnarray*}
\mathcal{P} &=&
  \bordermatrix{
  &\m w_{1}^{+}h_1 &\m w_{1}^{+}z_1 &\m w_{2}^{+}h_2 &\m w_{2}^{+}h_3 &\m w_{2}^{+}z_2 \cr\vbox{\hrule}
\m w_{1}^{+}h_1 & 2(\l_1+\l_3) & 0 & 2\l_3 & \l_4 & -2i\l_2 
   \cr
\m w_{1}^{+}z_1 & 0 & 2(\l_1+\l_3) & 2i\l_2 & 0 & 2\l_3  
   \cr
\m w_{2}^{+}h_2 & 2\l_3 & -2i\l_2 & 2(\l_1+\l_3) & -\l_4 & 0 
   \cr
\m w_{2}^{+}h_3 & \l_4 & 0 & -\l_4 & \l_5 & 0 
   \cr 
\m w_{2}^{+}z_2 & 2i\l_2 & 2\l_3 & 0 & 0 & 2(\l_1+\l_3)
   \cr
   }\,, \\
\mathcal{Q} &=&
  \bordermatrix{
  &\m w_{2}^{+}z_3 &\m w_{3}^{+}h_2 &\m w_{3}^{+}h_3 &\m w_{3}^{+}z_2 &\m w_{3}^{+}z_3 \cr\vbox{\hrule}
\m w_{1}^{+}h_1 & 0 & \l_4 & \frac{1}{2}(\l_6+2\l_7) & 0 & \frac{i}{2}(\l_6-2\l_7)
   \cr
\m w_{1}^{+}z_1 & \l_4 & 0 & \frac{-i}{2}(\l_6-2\l_7) & \l_4 & \frac{1}{2}(\l_6+2\l_7)
   \cr
\m w_{2}^{+}h_2 & 0 & -\l_4 & \frac{1}{2}(\l_6+2\l_7) & 0 & \frac{i}{2}(\l_6-2\l_7)
   \cr
\m w_{2}^{+}h_3 & 0 & \frac{1}{2}(\l_6+2\l_7) & 0 & \frac{-i}{2}(\l_6-2\l_7) & 0 
   \cr 
\m w_{2}^{+}z_2 & -\l_4 & 0 & \frac{-i}{2}(\l_6-2\l_7) & -\l_4 & \frac{1}{2}(\l_6+2\l_7)
   \cr
   }\,, \\
\mathcal{Q}^{\dagger} &=&
  \bordermatrix{
  &\m w_{1}^{+}h_1 &\m w_{1}^{+}z_1 &\m w_{2}^{+}h_2 &\m w_{2}^{+}h_3 &\m w_{2}^{+}z_2 \cr\vbox{\hrule}
\m w_{2}^{+}z_3 & 0 & \l_4 & 0 & 0 & -\l_4
   \cr
\m w_{3}^{+}h_2 & \l_4 & 0 & -\l_4 & \frac{1}{2}(\l_6+2\l_7) & 0
   \cr
\m w_{3}^{+}h_3 & \frac{1}{2}(\l_6+2\l_7) & \frac{i}{2}(\l_6-2\l_7) & \frac{1}{2}(\l_6+2\l_7) & 0 & \frac{i}{2}(\l_6-2\l_7)
   \cr
\m w_{3}^{+}z_2 & 0 & \l_4 & 0 & \frac{i}{2}(\l_6-2\l_7) & -\l_4
   \cr 
\m w_{3}^{+}z_3 & \frac{-i}{2}(\l_6-2\l_7) & \frac{1}{2}(\l_6+2\l_7) & \frac{-i}{2}(\l_6-2\l_7) & 0 & \frac{1}{2}(\l_6+2\l_7)
   \cr
   }\,, \\   
   \mbox{and,}\\        
\mathcal{R} &=&
  \bordermatrix{
  &\m w_{2}^{+}z_3 &\m w_{3}^{+}h_2 &\m w_{3}^{+}h_3 &\m w_{3}^{+}z_2 &\m w_{3}^{+}z_3 \cr\vbox{\hrule}
\m w_{2}^{+}z_3 & \l_5 & \frac{i}{2}(\l_6-2\l_7) & 0 & \frac{1}{2}(\l_6+2\l_7) & 0 
   \cr
\m w_{3}^{+}h_2 & \frac{-i}{2}(\l_6-2\l_7) & \l_5 & 0 & 0 & 0 
   \cr
\m w_{3}^{+}h_3 & 0 & 0 & 2\l_8 & 0 & 0
   \cr
\m w_{3}^{+}z_2 & \frac{1}{2}(\l_6+2\l_7) & 0 & 0 & \l_5 & 0 
   \cr 
\m w_{3}^{+}z_3 & 0 & 0 & 0 & 0 & 2\l_8
   \cr
   }\,.   
\end{eqnarray*} 
We find the eigenvalues of the matrix $\mathcal{K}$ to be $a_{i}^{\pm}~(i=1,\ldots,4),~b_{i}~(i=2,6)$ as listed in \Eqn{unitarity eq}. Thus we get all the eigenvalues of the matrix $\mathcal{M}_{CC}$.


\bibliographystyle{JHEP}
\bibliography{article.bib}

\providecommand{\href}[2]{#2}\begingroup\raggedright\begin{thebibliography}{10}

\bibitem{Aad:2012tfa}
{\bf ATLAS} Collaboration, G.~Aad et~al., {\it {Observation of a new particle
  in the search for the Standard Model Higgs boson with the ATLAS detector at
  the LHC}},  {\em Phys.Lett.} {\bf B716} (2012) 1--29,
  [\href{http://xxx.lanl.gov/abs/1207.7214}{{\tt arXiv:1207.7214}}].

\bibitem{Chatrchyan:2012ufa}
{\bf CMS} Collaboration, S.~Chatrchyan et~al., {\it {Observation of a new boson
  at a mass of 125 GeV with the CMS experiment at the LHC}},  {\em Phys.Lett.}
  {\bf B716} (2012) 30--61, [\href{http://xxx.lanl.gov/abs/1207.7235}{{\tt
  arXiv:1207.7235}}].

\bibitem{Kubo:2003pd}
J.~Kubo, {\it {Majorana phase in minimal S(3) invariant extension of the
  standard model}},  {\em Phys.Lett.} {\bf B578} (2004) 156--164,
  [\href{http://xxx.lanl.gov/abs/hep-ph/0309167}{{\tt hep-ph/0309167}}].

\bibitem{Koide:1999mx}
Y.~Koide, {\it {Universal seesaw mass matrix model with an S(3) symmetry}},
  {\em Phys.Rev.} {\bf D60} (1999) 077301,
  [\href{http://xxx.lanl.gov/abs/hep-ph/9905416}{{\tt hep-ph/9905416}}].

\bibitem{Harrison:2003aw}
P.~Harrison and W.~Scott, {\it {Permutation symmetry, tri - bimaximal neutrino
  mixing and the S3 group characters}},  {\em Phys.Lett.} {\bf B557} (2003) 76,
  [\href{http://xxx.lanl.gov/abs/hep-ph/0302025}{{\tt hep-ph/0302025}}].

\bibitem{Kubo:2003iw}
J.~Kubo, A.~Mondragon, M.~Mondragon, and E.~Rodriguez-Jauregui, {\it {The
  Flavor symmetry}},  {\em Prog.Theor.Phys.} {\bf 109} (2003) 795--807,
  [\href{http://xxx.lanl.gov/abs/hep-ph/0302196}{{\tt hep-ph/0302196}}].

\bibitem{Teshima:2005bk}
T.~Teshima, {\it {Flavor mass and mixing and S(3) symmetry: An S(3) invariant
  model reasonable to all}},  {\em Phys.Rev.} {\bf D73} (2006) 045019,
  [\href{http://xxx.lanl.gov/abs/hep-ph/0509094}{{\tt hep-ph/0509094}}].

\bibitem{Koide:2006vs}
Y.~Koide, {\it {S(3) symmetry and neutrino masses and mixings}},  {\em
  Eur.Phys.J.} {\bf C50} (2007) 809--816,
  [\href{http://xxx.lanl.gov/abs/hep-ph/0612058}{{\tt hep-ph/0612058}}].

\bibitem{Chen:2007zj}
C.-Y. Chen and L.~Wolfenstein, {\it {Consequences of approximate S(3) symmetry
  of the neutrino mass matrix}},  {\em Phys.Rev.} {\bf D77} (2008) 093009,
  [\href{http://xxx.lanl.gov/abs/0709.3767}{{\tt arXiv:0709.3767}}].

\bibitem{Mondragon:2007af}
A.~Mondragon, M.~Mondragon, and E.~Peinado, {\it {Lepton masses, mixings and
  FCNC in a minimal S(3)-invariant extension of the Standard Model}},  {\em
  Phys.Rev.} {\bf D76} (2007) 076003,
  [\href{http://xxx.lanl.gov/abs/0706.0354}{{\tt arXiv:0706.0354}}].

\bibitem{Jora:2009gz}
R.~Jora, J.~Schechter, and M.~Naeem~Shahid, {\it {Perturbed S(3) neutrinos}},
  {\em Phys.Rev.} {\bf D80} (2009) 093007,
  [\href{http://xxx.lanl.gov/abs/0909.4414}{{\tt arXiv:0909.4414}}].

\bibitem{Xing:2010iu}
Z.-z. Xing, D.~Yang, and S.~Zhou, {\it {Broken $S_3$ Flavor Symmetry of Leptons
  and Quarks: Mass Spectra and Flavor Mixing Patterns}},  {\em Phys.Lett.} {\bf
  B690} (2010) 304--310, [\href{http://xxx.lanl.gov/abs/1004.4234}{{\tt
  arXiv:1004.4234}}].

\bibitem{Kaneko:2010rx}
T.~Kaneko and H.~Sugawara, {\it {Broken $S_3$ Symmetry in Flavor Physics}},
  {\em Phys.Lett.} {\bf B697} (2011) 329--332,
  [\href{http://xxx.lanl.gov/abs/1011.5748}{{\tt arXiv:1011.5748}}].

\bibitem{Zhou:2011nu}
S.~Zhou, {\it {Relatively large theta13 and nearly maximal theta23 from the
  approximate S3 symmetry of lepton mass matrices}},  {\em Phys.Lett.} {\bf
  B704} (2011) 291--295, [\href{http://xxx.lanl.gov/abs/1106.4808}{{\tt
  arXiv:1106.4808}}].

\bibitem{Teshima:2011wg}
T.~Teshima and Y.~Okumura, {\it {Quark/lepton mass and mixing in $S_3$
  invariant model and CP-violation of neutrino}},  {\em Phys.Rev.} {\bf D84}
  (2011) 016003, [\href{http://xxx.lanl.gov/abs/1103.6127}{{\tt
  arXiv:1103.6127}}].

\bibitem{Dev:2011qy}
S.~Dev, S.~Gupta, and R.~R. Gautam, {\it {Broken $S_3$ Symmetry in the Neutrino
  Mass Matrix}},  {\em Phys.Lett.} {\bf B702} (2011) 28--33,
  [\href{http://xxx.lanl.gov/abs/1106.3873}{{\tt arXiv:1106.3873}}].

\bibitem{Dev:2012ns}
S.~Dev, R.~R. Gautam, and L.~Singh, {\it {Broken $S_3$ Symmetry in the Neutrino
  Mass Matrix and Non-Zero $\theta_{13}$}},  {\em Phys.Lett.} {\bf B708} (2012)
  284--289, [\href{http://xxx.lanl.gov/abs/1201.3755}{{\tt arXiv:1201.3755}}].

\bibitem{Meloni:2012ci}
D.~Meloni, {\it {$S_3$ as a flavour symmetry for quarks and leptons after the
  Daya Bay result on $\theta_{13}$}},  {\em JHEP} {\bf 1205} (2012) 124,
  [\href{http://xxx.lanl.gov/abs/1203.3126}{{\tt arXiv:1203.3126}}].

\bibitem{Dias:2012bh}
A.~Dias, A.~Machado, and C.~Nishi, {\it {An $S_3$ Model for Lepton Mass
  Matrices with Nearly Minimal Texture}},  {\em Phys.Rev.} {\bf D86} (2012)
  093005, [\href{http://xxx.lanl.gov/abs/1206.6362}{{\tt arXiv:1206.6362}}].

\bibitem{Siyeon:2012zu}
K.~Siyeon, {\it {Non-vanishing $U_{e3}$ under $S_3$ symmetry}},  {\em
  Eur.Phys.J.} {\bf 72} (2012) 2081,
  [\href{http://xxx.lanl.gov/abs/1203.1593}{{\tt arXiv:1203.1593}}].

\bibitem{Canales:2012dr}
F.~Gonzalez~Canales, A.~Mondragon, and M.~Mondragon, {\it {The $S_3$ Flavour
  Symmetry: Neutrino Masses and Mixings}},  {\em Fortsch.Phys.} {\bf 61} (2013)
  546--570, [\href{http://xxx.lanl.gov/abs/1205.4755}{{\tt arXiv:1205.4755}}].

\bibitem{Canales:2013cga}
F.~González~Canales, A.~Mondragón, M.~Mondragón, U.~J. Saldaña~Salazar, and
  L.~Velasco-Sevilla, {\it {Quark sector of S3 models: classification and
  comparison with experimental data}},  {\em Phys.Rev.} {\bf D88} (2013)
  096004, [\href{http://xxx.lanl.gov/abs/1304.6644}{{\tt arXiv:1304.6644}}].

\bibitem{Benaoum:2013ji}
H.~Benaoum, {\it {Broken $S_3$ Neutrinos}},  {\em Phys.Rev.} {\bf D87} (2013)
  073010, [\href{http://xxx.lanl.gov/abs/1302.0950}{{\tt arXiv:1302.0950}}].

\bibitem{Hernandez:2014vta}
A.~E.~C. Hernández, R.~Martinez, and J.~Nisperuza, {\it {$S_3$ flavour
  symmetry breaking scheme for understanding the quark mass and mixing pattern
  in $SU\left( 3\right) _{C}\otimes SU\left(3\right)_{L}\otimes U\left(
  1\right)_{X}$ models}},  \href{http://xxx.lanl.gov/abs/1401.0937}{{\tt
  arXiv:1401.0937}}.

\bibitem{Ma:2013zca}
E.~Ma and B.~Melic, {\it {Updated $S_{3}$ model of quarks}},  {\em Phys.Lett.}
  {\bf B725} (2013) 402--406, [\href{http://xxx.lanl.gov/abs/1303.6928}{{\tt
  arXiv:1303.6928}}].

\bibitem{Lee:1977eg}
B.~W. Lee, C.~Quigg, and H.~Thacker, {\it {Weak Interactions at Very
  High-Energies: The Role of the Higgs Boson Mass}},  {\em Phys.Rev.} {\bf D16}
  (1977) 1519.

\bibitem{Pakvasa:1977in}
S.~Pakvasa and H.~Sugawara, {\it {Discrete Symmetry and Cabibbo Angle}},  {\em
  Phys.Lett.} {\bf B73} (1978) 61.

\bibitem{Kubo:2004ps}
J.~Kubo, H.~Okada, and F.~Sakamaki, {\it {Higgs potential in minimal S(3)
  invariant extension of the standard model}},  {\em Phys.Rev.} {\bf D70}
  (2004) 036007, [\href{http://xxx.lanl.gov/abs/hep-ph/0402089}{{\tt
  hep-ph/0402089}}].

\bibitem{Koide:2005ep}
Y.~Koide, {\it {Permutation symmetry S(3) and VEV structure of flavor-triplet
  Higgs scalars}},  {\em Phys.Rev.} {\bf D73} (2006) 057901,
  [\href{http://xxx.lanl.gov/abs/hep-ph/0509214}{{\tt hep-ph/0509214}}].

\bibitem{Teshima:2012cg}
T.~Teshima, {\it {Higgs potential in $S_3$ invariant model for quark/lepton
  mass and mixing}},  {\em Phys.Rev.} {\bf D85} (2012) 105013,
  [\href{http://xxx.lanl.gov/abs/1202.4528}{{\tt arXiv:1202.4528}}].

\bibitem{Machado:2012ed}
A.~Machado and V.~Pleitez, {\it {Natural Flavour Conservation in a three
  Higgs-doublet Model}},  \href{http://xxx.lanl.gov/abs/1205.0995}{{\tt
  arXiv:1205.0995}}.

\bibitem{Gunion:2002zf}
J.~F. Gunion and H.~E. Haber, {\it {The CP conserving two Higgs doublet model:
  The Approach to the decoupling limit}},  {\em Phys.Rev.} {\bf D67} (2003)
  075019, [\href{http://xxx.lanl.gov/abs/hep-ph/0207010}{{\tt
  hep-ph/0207010}}].

\bibitem{Bhattacharyya:2010hp}
G.~Bhattacharyya, P.~Leser, and H.~Pas, {\it {Exotic Higgs boson decay modes as
  a harbinger of $S_3$ flavor symmetry}},  {\em Phys.Rev.} {\bf D83} (2011)
  011701, [\href{http://xxx.lanl.gov/abs/1006.5597}{{\tt arXiv:1006.5597}}].

\bibitem{Bhattacharyya:2012ze}
G.~Bhattacharyya, P.~Leser, and H.~Pas, {\it {Novel signatures of the Higgs
  sector from S3 flavor symmetry}},  {\em Phys.Rev.} {\bf D86} (2012) 036009,
  [\href{http://xxx.lanl.gov/abs/1206.4202}{{\tt arXiv:1206.4202}}].

\bibitem{Chen:2004rr}
S.-L. Chen, M.~Frigerio, and E.~Ma, {\it {Large neutrino mixing and normal mass
  hierarchy: A Discrete understanding}},  {\em Phys.Rev.} {\bf D70} (2004)
  073008, [\href{http://xxx.lanl.gov/abs/hep-ph/0404084}{{\tt
  hep-ph/0404084}}].

\bibitem{Beltran:2009zz}
O.~F. Beltran, M.~Mondragon, and E.~Rodriguez-Jauregui, {\it {Conditions for
  vacuum stability in an S(3) extension of the standard model}},  {\em
  J.Phys.Conf.Ser.} {\bf 171} (2009) 012028.

\bibitem{Barradas-Guevara:2014yoa}
E.~Barradas-Guevara, O.~Felix-Beltran, and E.~R. Jauregui, {\it {S(3) flavoured
  Higgs model trilinear self-couplings}},
  \href{http://xxx.lanl.gov/abs/1402.2244}{{\tt arXiv:1402.2244}}.

\bibitem{Maalampi:1991fb}
J.~Maalampi, J.~Sirkka, and I.~Vilja, {\it {Tree level unitarity and triviality
  bounds for two Higgs models}},  {\em Phys.Lett.} {\bf B265} (1991) 371--376.

\bibitem{Kanemura:1993hm}
S.~Kanemura, T.~Kubota, and E.~Takasugi, {\it {Lee-Quigg-Thacker bounds for
  Higgs boson masses in a two doublet model}},  {\em Phys.Lett.} {\bf B313}
  (1993) 155--160, [\href{http://xxx.lanl.gov/abs/hep-ph/9303263}{{\tt
  hep-ph/9303263}}].

\bibitem{Akeroyd:2000wc}
A.~G. Akeroyd, A.~Arhrib, and E.-M. Naimi, {\it {Note on tree level unitarity
  in the general two Higgs doublet model}},  {\em Phys.Lett.} {\bf B490} (2000)
  119--124, [\href{http://xxx.lanl.gov/abs/hep-ph/0006035}{{\tt
  hep-ph/0006035}}].

\bibitem{Horejsi:2005da}
J.~Horejsi and M.~Kladiva, {\it {Tree-unitarity bounds for THDM Higgs masses
  revisited}},  {\em Eur.Phys.J.} {\bf C46} (2006) 81--91,
  [\href{http://xxx.lanl.gov/abs/hep-ph/0510154}{{\tt hep-ph/0510154}}].

\bibitem{Pal:1994jk}
P.~B. Pal, {\it {What is the equivalence theorem really?}},
  \href{http://xxx.lanl.gov/abs/hep-ph/9405362}{{\tt hep-ph/9405362}}.

\bibitem{Horejsi:1995jj}
J.~Horejsi, {\it {Electroweak interactions and high-energy limit: An
  Introduction to equivalence theorem}},  {\em Czech.J.Phys.} {\bf 47} (1997)
  951--977, [\href{http://xxx.lanl.gov/abs/hep-ph/9603321}{{\tt
  hep-ph/9603321}}].

\bibitem{Searches:2001ac}
{\bf LEP Higgs Working Group for Higgs boson searches, ALEPH Collaboration,
  DELPHI Collaboration, L3 Collaboration, OPAL Collaboration} Collaboration,
  {\it {Search for charged Higgs bosons: Preliminary combined results using LEP
  data collected at energies up to 209-GeV}},
  \href{http://xxx.lanl.gov/abs/hep-ex/0107031}{{\tt hep-ex/0107031}}.

\bibitem{Gunion:1989we}
J.~F. Gunion, H.~E. Haber, G.~L. Kane, and S.~Dawson, {\it {The Higgs Hunter's
  Guide}},  {\em Front.Phys.} {\bf 80} (2000) 1--448.

\bibitem{Djouadi:1996yq}
A.~Djouadi, V.~Driesen, W.~Hollik, and A.~Kraft, {\it {The Higgs photon - Z
  boson coupling revisited}},  {\em Eur.Phys.J.} {\bf C1} 163--175.

\bibitem{Arhrib:2003ph}
A.~Arhrib, M.~Capdequi~Peyranere, W.~Hollik, and S.~Penaranda, {\it {Higgs
  decays in the two Higgs doublet model: Large quantum effects in the
  decoupling regime}},  {\em Phys.Lett.} {\bf B579} (2004) 361--370,
  [\href{http://xxx.lanl.gov/abs/hep-ph/0307391}{{\tt hep-ph/0307391}}].

\bibitem{Bhattacharyya:2013rya}
G.~Bhattacharyya, D.~Das, P.~B. Pal, and M.~Rebelo, {\it {Scalar sector
  properties of two-Higgs-doublet models with a global U(1) symmetry}},  {\em
  JHEP} {\bf 1310} (2013) 081, [\href{http://xxx.lanl.gov/abs/1308.4297}{{\tt
  arXiv:1308.4297}}].

\bibitem{Ferreira:2014naa}
P.~Ferreira, J.~F. Gunion, H.~E. Haber, and R.~Santos, {\it {Probing wrong-sign
  Yukawa couplings at the LHC and a future linear collider}},
  \href{http://xxx.lanl.gov/abs/1403.4736}{{\tt arXiv:1403.4736}}.

\bibitem{pdg}
{M. Carena, {\em et al.} (Particle Data Group)}, {\it Status of higgs boson
  physics},  Nov., 2013.
\newblock \url{http://pdg.lbl.gov/2013/reviews/rpp2013-rev-higgs-boson.pdf}.

\end{thebibliography}\endgroup
\end{document}